%% file: amy_notes.tex
\documentclass{article}

\usepackage{arxiv}

\usepackage{appendix}
\appendixtitleon
\appendixtitletocon

\usepackage[utf8]{inputenc} 
\usepackage[T1]{fontenc}    
\usepackage[colorlinks=true,linkcolor=magenta,citecolor=blue,pagebackref]{hyperref}
\usepackage{url}            
\usepackage{booktabs}       
\usepackage{amsfonts}       
\usepackage{amsmath}
\usepackage{nicefrac}       

\usepackage{natbib}
\usepackage{doi}

\usepackage{subfigure, epsfig}

\bibpunct[, ]{(}{)}{,}{a}{}{,}%

\newtheorem{theorem}{Theorem}
\newtheorem{lemma}{Lemma}
\newtheorem{proposition}{Proposition}
\newtheorem{corollary}{Corollary}

\newtheorem{conjecture}{Conjecture}
\newtheorem{hypothesis}{Hypothesis}
\newtheorem{assumption}{Assumption}
\newtheorem{remark}{Remark}
\newtheorem{example}{Example}

\newtheorem{definition}{Definition}

\newtheorem{appthm}{Theorem}

\newtheorem{appassm}{Assumption}

\newtheorem{appdefn}{Definition}

\newtheorem{applem}{Lemma}

\usepackage{times}
\usepackage{amssymb}
\usepackage{latexsym}
\usepackage{graphicx}
\usepackage{color}
\usepackage{enumitem}

\usepackage{multirow}
\usepackage{threeparttable}

\usepackage[labelfont=bf]{caption}
\usepackage{soul}

\usepackage{longtable}
\usepackage{makecell}
\usepackage{multirow}
\usepackage{float}
\usepackage{bbm}

\usepackage{svg}

\usepackage{bibentry}
\usepackage{fancyhdr}
\usepackage{booktabs}
\usepackage{titlesec}
\usepackage{url}

\usepackage{pgf,tikz}
\usetikzlibrary{arrows,shapes.arrows,shapes.geometric,shapes.multipart,
decorations.pathmorphing,positioning,swigs}
\usetikzlibrary{decorations.pathreplacing}
\usetikzlibrary{calc,arrows.meta,positioning}

\pgfkeys{/tikz/swig vsplit/.code={%
              \pgfkeys{/tikz/swig vsplit/.cd,#1}%
               }}%
\pgfkeys{/tikz/swig vsplit/gap/.initial = 0.8em}%
\pgfkeys{/tikz/swig vsplit/line width left/.initial = \the\pgflinewidth}%
\pgfkeys{/tikz/swig vsplit/line width right/.initial = \the\pgflinewidth}%
\pgfkeys{/tikz/swig vsplit/inner line width left/.initial = 0pt}%
\pgfkeys{/tikz/swig vsplit/inner line width right/.initial = 0pt}%

\usepackage{scalefnt}

\usetikzlibrary{shapes,decorations,arrows,calc,arrows.meta,fit,positioning}
\tikzset{
    -Latex,auto,node distance =1 cm and 1 cm,semithick,
    state/.style ={circle, draw, minimum width = 0.5 cm},
    point/.style = {circle, draw, inner sep=0.04cm,fill,node contents={}},
    bidirected/.style={Latex-Latex,dashed},
    el/.style = {inner sep=2pt, align=left, sloped}
}

\usepackage{listings}

\definecolor{codegreen}{rgb}{0,0.6,0}
\definecolor{codegray}{rgb}{0.5,0.5,0.5}
\definecolor{codepurple}{rgb}{0.58,0,0.82}
\definecolor{backcolour}{rgb}{0.95,0.95,0.92}

\lstdefinestyle{mystyle}{
    backgroundcolor=\color{backcolour},   
    commentstyle=\color{codegreen},
    keywordstyle=\color{magenta},
    numberstyle=\tiny\color{codegray},
    stringstyle=\color{codepurple},
    basicstyle=\ttfamily\footnotesize,
    breakatwhitespace=false,         
    breaklines=true,                 
    captionpos=b,                    
    keepspaces=true,                 
    numbers=left,                    
    numbersep=5pt,                  
    showspaces=false,                
    showstringspaces=false,
    showtabs=false,                  
    tabsize=2
}

\newcommand{\ind}{{\perp \!\!\! \perp}}

\newcommand{\ls}[1]
  {\dimen0=\fontdimen6\the\font \lineskip=#1\dimen0
  \advance\lineskip.5\fontdimen5\the\font \advance\lineskip-\dimen0
  \lineskiplimit=.9\lineskip \baselineskip=\lineskip
  \advance\baselineskip\dimen0 \normallineskip\lineskip
  \normallineskiplimit\lineskiplimit \normalbaselineskip\baselineskip
  \ignorespaces }

\newcommand{\Pf}{\paragraph{{\bf Proof.}}}       
\newcommand{\blot}{\hfill{\vrule height .9ex width .8ex depth -.1ex }}
\newcommand{\EndPf}{\hfill $\blot$ \medskip}     

\iffalse 
\usepackage{amsfonts}
\usepackage{verbatim}
\usepackage{bbm}

\usepackage{graphicx}

\usepackage{authblk}

\usepackage{amsthm}
\newcommand{\field}[1]{\mathbb{#1}}
\DeclareMathOperator{\PR}{\field{P}}             
\DeclareMathOperator{\E}{\field{E}}              
\def\N{\field{N}}                                
\def\R{\field{R}}                                
\def\F{\field{F}}                                
\def\A{\field{A}}                                

\def\Pa{\mathop{{\rm{Pa}}}\nolimits}  

\else

\def\Pa{\mathop{{\rm{Pa}}}\nolimits}  

\def\PR{\mathop{\rm I\kern -0.20em P}\nolimits}  
\def\E{\mathop{\rm I\kern -0.20em E}\nolimits}   
\def\N{\mathop{\rm I\kern -0.20em N}\nolimits}   
\def\R{\mathop{\rm I\kern -0.20em R}\nolimits}   
\def\F{\mathop{\rm I\kern -0.20em F}\nolimits}   
\def\A{{A}}                                      
\fi

\graphicspath{{./}{Figs/}}

\usepackage{longtable}

\date{\small \textit{\today}}

\newcommand{\sref}[1]{\ref{#1}}
\newcommand{\mref}[1]{\ref{#1}}

\title{Unveiling Bias in Sequential Decision Making: \\A Causal Inference Approach for Stochastic Service Systems}

\author{ \hspace{1mm}Juan Camilo David\\
	Department of Industrial and Systems Engineering\\
	University of Wisconsin-Madison\\
	Madison, WI \\
	\texttt{\href{mailto:@wisc.ed}{davidgomez@wisc.edu}} \\
	\And
	\hspace{1mm}Amy Cochran \\
	Department of Population Health Sciences \\ and Department of Mathematics\\
	University of Wisconsin-Madison\\
	Madison, WI \\
	\texttt{\href{mailto:@wisc.ed}{cochran4@wisc.edu}} \\
	\And
	\hspace{1mm}Gabriel Zayas-Cab\'{a}n \\
	Department of Industrial and Systems Engineering\\
	University of Wisconsin-Madison\\
	Madison, WI \\
	\texttt{\href{mailto:zayascaban@wisc.edu}{zayascaban@wisc.edu}} \\
}

\renewcommand{\shorttitle}{Unveiling Bias in Sequential Decision Making}

\hypersetup{
pdftitle={Sequential Bias},
pdfsubject={med},
pdfauthor={David Juan, Cochran Amy, Zayas-Caban Gabriel},
pdfkeywords={Bias, GEE},
}

\begin{document}

\maketitle

\begin{abstract}
In many stochastic service systems, decision-makers find themselves making a sequence of decisions, with the number of decisions being unpredictable. To enhance these decisions, it is crucial to uncover the causal impact these decisions have through careful analysis of observational data from the system. However, these decisions are not made independently, as they are shaped by previous decisions and outcomes. This phenomenon is called \emph{sequential bias} and violates a key assumption in causal inference that one person's decision does not interfere with the potential outcomes of another. To address this issue, we establish a connection between sequential bias and the subfield of causal inference known as dynamic treatment regimes. We expand these frameworks to account for the random number of decisions by modeling the decision-making process as a marked point process. Consequently, we can define and identify causal effects to quantify sequential bias. Moreover, we propose estimators and explore their properties, including double robustness and semiparametric efficiency. In a case study of 27,831 encounters with a large academic emergency department, we use our approach to demonstrate that the decision to route a patient to an area for low acuity patients has a significant impact on the care of future patients.
\end{abstract}


\section{Introduction}

Sequential decision-making problems are pervasive and thus the subject of extensive analysis in Operations Research and Management Sciences, Computer Science, Statistics, and other disciplines. These problems are primarily motivated by the necessity of enhancing a sequence of decisions for a \emph{single} unit, such as a person or single job. One example is medical decision-making, which searches for strategies that optimize outcomes for an individual patient. This is often achieved using Markov decision processes \citep{alagoz2010markov,steimle2017markov}. Another example is dynamic treatment regimes (DTRs), which aims to estimate the causal impact of sequences of treatment decisions compared to a baseline strategy. These estimates are then used to find a sequence of decisions that enhances outcomes for each patient. This approach has considerable attention in the statistical literature \citep{robins_causal_1997,susan2003,chakraborty2013statistical,tsiatis2019dynamic}. 

Instead of examining multiple decisions for a single unit, we explore a scenario where a single decision-maker navigates a sequence of decisions over time for \emph{multiple} units. Our objective is not to identify optimal strategies for such scenarios, but rather to assess and estimate the direct or causal influence that one unit's decision has on future unit's decisions and outcomes. Inspired by the psychology literature, we refer to this influence as \emph{sequential bias} \citep{yu_sequential_2008}.

Service systems have provided many compelling examples of sequential bias. The authors' inspiration derives from their research of split-flow models in emergency departments (EDs), where the triage nurse is replaced by a physician  \citep{zayas2016dynamic,gomez2022evaluation}. As part of split flow, the physician at triage decides between routing patients to a vertical area for low-acuity patients or a traditional room. In resource-constrained settings like the ED, physicians must weigh the care of the current patient against the needs of future patients. This prompts the question of whether a physician's routing decision for one patient influences the care of future patients. 

Another example, with which researchers studying service systems are likely familiar, is the organ transplantation process. Led by the Organ Procurement and Transplantation Network, the transplantation process involves a list of patients awaiting organs, prioritized based on availability, type, location, and health status. When an organ becomes available, potential recipients are identified according to compatibility and urgency. A patient's transplant team, acting as decision-makers, assess organ compatibility and faces a time-sensitive binary choice of accepting or rejecting the organ. Due to the limited supply of organs, this sequential decision-making process is susceptible to sequential bias, wherein the current decision carries life-and-death consequences for future patients.

Sequential bias is relevant in other domains.  In many selection processes, for example, evaluators assess candidates in a sequential manner to determine their suitability for a role, position, or opportunity. These processes can include grading students, judging sporting competitions, determining criminal sentencing, or approving loans, among others. The prior evaluations of candidates can impact subsequent scores assigned by evaluators, potentially introducing bias. The order in which candidates are evaluated plays a crucial role as it can affect the outcomes and favor certain individuals while disadvantaging others \citep{chen_decision_2016, goldbach_sequential_2022}. 

A common thread runs through all these examples, allowing for a single conceptual framework.  A decision-maker encounters various jobs within a stochastic service system and must make decisions for each job sequentially over time. The number of jobs being intervened upon is random and may be beyond the immediate control of the decision-maker. Each job shares a common set of binary intervention options, and the decision for each job bears consequences for the job itself. These decisions are determined based on contextual information and the history of prior interventions, contexts, and outcomes. Additionally, the decision-maker lacks information on future contexts or outcomes when making the current decision. The objective is to quantify sequential bias using observational data from the system. This entails defining, identifying, and estimating the causal effect that a decision for the current job has on the decisions and outcomes of future jobs. By accurately quantifying this bias, we unlock opportunities to improve decision-making in stochastic service systems through stochastic modeling and optimization.

Performing causal inference within this context presents two issues. First, there is the issue of interference, where the decision for one job influences decisions and outcomes of future jobs \citep{cox_planning_1958}. Interference violates a core assumption in causal inference known as the Stable Unit Treatment Value Assumption (SUTVA) \citep{rubin1980randomization}. An immediate implication is that the potential decisions and outcomes of a future job is dependent on the entire history of past interventions. The collection of these histories grows exponentially in the number of jobs, making it impractical to contrast all possible sequences of interventions. The DTR literature also encounters this exponential growth, offering us ideas for focusing on specific causal contrasts \citep{guo2021discussion}. Causal blip and excursion effects are examples of such focused approaches, as they provide more manageable linear growth in the number of decisions \citep{robins_causal_1997,boruvka_assessing_2018,qian_estimating_2021}. 

Second, there is the issue of a random number of variables, further complicated by the possible dependence of this random number on the history of prior interventions, contexts, and outcomes. This brings into question on how causal frameworks, such as Pearl's do-calculus \citep{pearl_causality_2000}, Neyman-Rubin potential outcomes \citep{neyman1923application,rubin1974estimating}, or specific frameworks for DTRs \citep{susan2003}, can be readily applied. These frameworks are typically intended for scenarios where there are a fixed and finite number of random variables. One option would be to set an upper bound on the random number of jobs. However, it is often preferred to avoid such a restrictive assumption in stochastic service systems. Consequently, the random number of jobs leads us to, in effect, deal with an infinite number of random variables, presenting challenges across various aspects. These challenges include ensuring the existence of probability distributions, handling integrability, and carefully managing the interchangeability of summation and differentiation in this infinite context.

In this paper, we contribute a solution to these issues as follows. We define a causal model to capture the sequential decision-making scenario and the random number of jobs. To address the varying number of variables, we artificially extend the variables to represent a marked point process \citep{jacobsen2006point}. This extension results in an infinite number of random variables. Subsequently, we utilize Markov kernels to construct a probability distribution for the marked point process and to represent hypothetical scenarios were decisions to be modified. This construction aligns with the causal frameworks proposed by \citet{richardson2013single} and \citet{malinsky2019potential}, which unified Pearl's do-calculus \citep{pearl_causality_2000} with the Neyman-Rubin potential outcomes framework. It also facilitates the use of the single-world intervention graph (SWIG) and encompasses common causal inference assumptions such as consistency and sequential ignorability. For clarity, we illustrate the causal model with directed acyclic graphs restricted to a finite number of variables.

The subsequent step is defining and identifying causal effects. These causal effects are defined in a manner that closely resembles the causal contrasts utilized by \citet{boruvka_assessing_2018} and \citet{qian_estimating_2021}, while also accounting for the adjustments needed to accommodate the random number of original variables. We introduce two types of causal effects: the lag effect and the marginalized lag effect, where the latter is a dimensionally reduced version of the former. A crucial consideration for these effects is the need to condition on post-intervention variables. This is to ensure that we exclusively measure the causal impact on future jobs that actually exist, given that the number of jobs is random. For insightful discussion on conditioning on post-intervention variables, refer to \citet{pearl2015conditioning}. Importantly, we demonstrate that our causal effects can be nonparametrically identified.

The next step is estimation, for which we propose an estimating equation approach with careful consideration for integrability. We prove our estimator is doubly-robust and apply standard statistical arguments to establish its consistency and asymptotic normality \citep{van2000asymptotic}. Furthermore, using techniques from semiparametric theory \citep{tsiatis_semiparametric_2006}, we derive the efficient score for a specific case of our lag effect, which supports our choice of estimator. Last, we apply our estimator to electronic healthcare records (n=27,831) from a large academic hospital that operates a split-flow model in the ED. We investigate the causal impact that routing the current patient to a vertical area over a traditional room has on the subsequent patient. 

The remainder of this article is structured as follows. Section~\ref{sec:lit_review} summarizes relevant studies from the literature, encompassing topics such as DTRs and sequential randomized trials, interference in causal inference, and causal reasoning for stochastic processes. It also highlights the contribution of this study in relation to the surveyed literature. Section~\ref{sec:bias_framework} introduces the causal model employed in this paper, utilizing marked point processes, and defines the causal effects of interest. Additionally, it presents conditions under which these causal effects can be identified from observational data. Moving forward, Section~\ref{sec:estimation} presents an estimator for these causal effects, accompanied by derivations of its asymptotic properties such as double robustness and its efficient score. Section~\ref{sec:bias_case_study} applies the estimation procedure to the split-flow example discussed earlier, aiming to determine the impact of prior routing decisions on the subsequent patient's outcomes. Section~\ref{sec:bias_discussion} concludes our study.

\section{Literature Review}\label{sec:lit_review}

\subsection{Dynamic Treatment Regimes and Sequential Randomized Trials} 

The current study relates to the subfield of causal inference that explores sequential decision-making for individual units. Specifically, it closely aligns with DTRs, sequential multiple assignment randomized trials (SMARTs), and their high-frequency counterparts known as micro-randomized trials.

A DTR is a specified sequence of treatment decision rules that determines how to adapt the delivery of treatments over time in response to an individual's changing health and other time-varying contextual factors. A DTR typically consists of multiple stages, where each stage considers an individual's medical history and current health information to recommend the next treatment. Extensive research has focused on determining the optimal DTR \citep{susan2003, laber2014dynamic,li2023optimal} and estimating the causal effects of specific treatment regimes relative to an alternative treatment regime~\citep{robins1986new,robins_causal_1997, susan2003, chakraborty_dynamic_2014}. One of the key challenges addressed in this literature is the exponential growth in the total number of treatment regimes as the number of treatment decisions increases. For instance, binary decisions at four stages results in a total of $2^4$ possible treatment sequences. As a consequence, the amount of available data for comparing two treatment sequences significantly decreases as more decision stages are added. This reduction in data introduces greater uncertainty in the comparisons and optimization of DTRs. 

When faced with an overwhelming number of treatment sequences, a commonly employed approach is to focus on DTRs that can be represented using observed treatment decisions. These DTRs are typically characterized by a linear increase in their complexity with respect to treatment stages. These concepts are captured in modeling frameworks like the structural nested mean model \citep{robins_correcting_1994} and quantified through measures such as \textit{blip effects}~\citep{robins_causal_1997,wang_new_2020} and \textit{causal excursion effects} \cite{qian_estimating_2021,shi2022estimating}, among others \citep{guo2021discussion}. To illustrate, a blip effect can compare treatment regimes that initially align with the observed treatment until a certain stage, after which each regime discontinues the treatment. The sole distinction between these regimes lies in the timing of when the treatment is discontinued. In general, these causal effects are typically described as the average response of an individual at a specific stage, considering two different treatment regimes, while conditioning on their previous treatment history and time-dependent factors. The DTR literature employs standard assumptions for causal inference, such as positivity, consistency, and conditional exchangeability. These assumptions play a crucial role in non-parametrically identifying the causal effects and enable us to perform causal inference even in non-experimental settings, such as when working with observational data. 

Related to DTRs are experimental designs known as SMARTs and also referred to as \textit{alternative designed randomized trials} by \citet{robins1986new}. These designs involve repeatedly assigning participants to different treatment conditions at multiple stages, considering their individual characteristics and the historical information of covariates and treatments \citep{philip2000, lavori2004dynamic}. These trials help inform DTRs, providing at evidence of a randomized control trial with the advantage of tailoring treatment decisions more efficiently based on a patient's current and past context. For example,  if a patient does not respond to a treatment, they can be re-randomized to alternative treatments in order to find a more effective approach. The seminal work of \citet{murphy2005experimental} presents a robust framework for designing SMARTs. Moreover, the statistical methods of a SMART can be adapted to non-experimental contexts with a few key distinctions. Specifically, assumptions regarding certain conditional independence relations are necessary, and the randomization probabilities of treatment assignments typically need to be estimated from data. To safeguard inferences against incorrect assumptions, robust methods are recommended \citep{guo2021discussion}.

Similar to SMARTs, micro-randomized trials involve repeatedly randomizing participants to different treatment conditions over time \citep{klasnja_micro-randomized_2015}. These trials build on the causal inference concepts for analyzing DTRs from longitudinal data,  introduced in~\citet{robins_causal_1997} and \citet{gill_causal_2001}. In a micro-randomized trial, randomization occurs frequently, often multiple times a day. This high-frequency randomization allows researchers to adapt interventions to an individual's rapidly changing context, such as variations in stress or mood. Micro-randomized trials are especially suitable for evaluating mobile health interventions \citep{klasnja_micro-randomized_2015, boruvka_assessing_2018, qian_estimating_2021} and developing just-in-time adaptive interventions \citep{nahum-shani_just--time_2017}. As an example, an author of this study conducted micro-randomized trials to evaluate a mobile version of a therapy called acceptance and commitment therapy \citep{cochran2023mobile,thomas2023mobile}. 
Extensive efforts have been made to devise statistical methods for micro-randomized trials \citep{klasnja_micro-randomized_2015,boruvka_assessing_2018,qian_estimating_2021}.

When designing DTRs, SMARTs, and micro-randomized trials, decision epochs represent predetermined points in time when treatment decisions are made for each individual. Notably, decision epochs have a finite and fixed number in these designs. For example, a SMART frequently uses two epochs: a trial's onset and a later time when treatment response is evaluated. Meanwhile, micro-randomized trials incorporate multiple decision epochs throughout each day over a specified duration, such as fixed morning and evening time windows over six weeks in the aforementioned trials \citep{cochran2023mobile,thomas2023mobile}. Consequently, existing statistical methods have historically overlooked the added complexity of a random number of decision epochs. 

By contrast, stochastic service systems encounter a random number of decision epochs when performing sequential decision-making. This variability can arise from the unpredictable number of arrivals to the service system and the random duration required to service jobs within the system. In a stochastic service system, the number of decision epochs is not only random but can also be influenced by the system's history. Decisions may be restricted to specific time periods (e.g., 9am to 5pm) or restricted to a fixed number of jobs receiving a higher level of resources. Due to the system's stochastic nature, these constraints result in a random number of decisions.

\subsection{Interference in Causal Inference}

A stochastic service system involves a sequence of decisions for different jobs or individuals, which differs from DTRs, SMARTs, and micro-randomized trials where sequential decisions are often centered around the same individual.  This presents a challenge because most literature on causal inference relies on SUTVA \citep{rubin1980randomization, rubin_causal_2005}. Part of SUTVA is the no interference assumption~\citep{cox_planning_1958}, which says that the potential outcomes of a specific unit are unaffected by the decisions on other units. The sequential decision processes analyzed in this work violate the no interference assumption, as the decisions made for previous jobs can impact the potential outcomes of future jobs. Consequently, we must appropriately account for interference when identifying and estimating causal effects in the context of sequential decision-making within a stochastic service system.

Early approaches to interference involved dividing units into equal-sized and non-overlapping blocks, allowing interference within blocks but not across them, known as partial interference~\citep{hong_evaluating_2006, sobel_what_2006, hudgens_toward_2008}. Recent advancements have allowed for more arbitrary patterns of interference \citep{verbitsky-savitz_causal_2012, sofrygin_semi-parametric_2017}. These approaches permit interference among subsets of units based on spatial proximity, social ties, and other measures of proximity. An example of modeling general interference without partial interference is presented by \citet{aronow_estimating_2017}, where a function is introduced to model exposure using aggregated data on the number of exposed neighbors. This approach is flexible as it allows an arbitrary number of neighbors. However, it has the limitation of not allowing other types of interactions, such as contagion, defined as the outcome of a unit causing the outcome of a different unit. More recently, ~\citet{tchetgen_tchetgen_auto-g-computation_2021} proposed a fairly flexible method for inferring causal inferences from complex networks, and~\citet{zhang_causal_nodate} build a model where specific forms of interactions (e.g., spillovers, contagion) can be explicitly incorporated. The study by~\citet{zhang_causal_nodate} also develops an algorithm to quantify interaction bias and conditions under which the bias may be ignored. However, the framework is restricted to causal linear models.

Our view is that interference within a wide range of stochastic service systems is notably less complex compared to the arbitrary interference highlighted earlier. There are two reasons for this. First, stochastic service systems have a sequential nature, meaning that future events cannot causally influence past events. This sequentiality leads to a progressive accumulation of information over time, akin to a filtration in the context of a stochastic process. Interestingly, this sequential interference gives rise to similar relationships of conditional independence between observed variables as seen in DTRs, SMARTs, and micro-randomized trials. As a result, the approaches used to handle conditional independence in the latter can be applied to address interference in stochastic service systems.

Second, we can anticipate partial interference in numerous real-world service systems, which can be attributed to two underlying mechanisms. One mechanism is the operation of service systems within fixed time windows, such as 9am to 5pm. This holds for the ED split-flow model that served as the inspiration for our research. It is reasonable to assume that interference occurs among jobs within the same time window but not between jobs across different time windows. A second mechanism is the use of distinct decision-makers in a service system. There are situations where it is justifiable to assume that interference occurs among jobs handled by the same decision-maker but not between jobs assigned to different decision-makers. Thus, the concept of partial interference, as discussed in previous studies \citep{hong_evaluating_2006, sobel_what_2006, hudgens_toward_2008}, can be applied. Specifically, we adopt an assumption of independent data panels, whereby interference occurs between jobs within each panel, but not between jobs from different panels.

\subsection{Stochastic Processes and Causal Inference}

Existing literature on causal inference for stochastic systems  remains relatively scarce, but is increasingly gaining recognition \citep{didelez2008graphical,33454a05-596e-3e2d-a725-b726efc21380,roysland2012counterfactual,didelez2015causal,gao2021causal}. Significantly, many foundational elements commonly employed in causal inference in other contexts, such as causal models and graphical representations, face significant challenges when applied to stochastic systems. Early representative work is the study by \citet{didelez2008graphical} which proposes a new class of graphical models capturing the dependence structure of events that occur in time via marked point processes. 
Although our research does not directly employ these graphical models, their work inspired our adoption of marked point processes. Another distinction is our interest in identification and estimation of causal effects using observational data, as opposed to graphical analyses.

\section{Causal Framework}\label{sec:bias_framework}

\subsection{Notation}

We use the following notation:
\begin{itemize}
\item Capital letters for random variables (e.g., $Z$).
\item Greek letters for parameters (e.g., $\theta, \xi, \eta, \alpha,\beta$).
\item ${\cal X}^{(Z)}$ for the target space of a random variable $Z$. 
\item Lower case letters for observations of a random variable (e.g., $z \in {\cal X}^{(Z)}$).
\item Subscripts for vectors or sequences, e.g., $Z_{1:k}$ is $(Z_1, \ldots, Z_k)$ and $Z_{B}$ is  $(Z_{i_1}, \ldots, Z_{i_k})$ when $B=\{i_1,\ldots,i_k\} \subseteq \N$.
\item Apostrophe for transpose, e.g. $Z'$ is the transpose of $Z$.
\item $\| Z\|$ for the $L_2$-norm of $Z$.
\item $\nabla_{\alpha}$ for partial differentiation with respect to $\alpha$. 
\item $\PR^{(Z)}$ for a Markov kernel from some source ${\cal X}^{(W)}$ to a target ${\cal X}^{(Z)}$, which are regular conditional distributions that generalize the transition matrix for Markov chains to general state spaces.
\end{itemize}
We adopt the convention that any such Markov kernel is a probability distribution on ${\cal X}^{(Z)}$ when the source ${\cal X}^{(W)}$ is $\emptyset$ and can be viewed as a Markov kernel with source ${\cal X}^{(W,V)}$ that is insensitive to some of its arguments, viz,
$$\PR^{(Z)}( \cdot | w, v) = \PR^{(Z)}( \cdot | w ), \quad \forall\, (w,v) \in {\cal X}^{(W,V)}.$$ 
In addition, we adopt the convention that $Z_{1:k}$ and ${\cal X}^{Z_{1:k}}$ are empty when $k=0$ and that ${\cal X}^{(Z_{1:2})}$ represents the product space ${\cal X}^{(Z_1)} \times {\cal X}^{(Z_2)}$ associated with $Z_{1:2}$ (similarly for sequences). 

\subsection{Setting}\label{sec:setting}

In our stochastic service system, jobs (such as customers, patients, or units) are assumed to arrive one at a time and are processed individually. Each job is associated with specific characteristics like age, sex, race, chief complaint, comorbidities, and the current capacity of the system. A decision-maker makes a binary decision for each job, which can be influenced by the job's characteristics and by information on jobs in the past. The decision leads to an outcome for the job, such as the length of time they stay in the system or the number of tests performed. In this system, the decision-maker processes a \emph{random} number of jobs in any given shift or day, which is called a panel. It is assumed that multiple panels are associated with the service system.

We collect data on a panel using several variables. Let $K \geq 1$ be the random number of jobs in a panel, and let $k$ denote the index of a job within a panel ($k=1,\ldots,K$). Each job $k$ is associated with three random variables: 
\begin{itemize}
    \item $X_k \in \mathbb{R}^l$ (a vector of characteristics associated with the job), 
    \item $A_k \in \{0,1\}$ (a binary intervention), and 
    \item $Y_k \in \mathbb{R}$ (an outcome of interest). 
\end{itemize}
We assume that the set of variables from a single panel are mutually independent across panels. The history of all three variables up to and including job $k$ is denoted by $H_k = (X_{1:k},A_{1:k},Y_{1:k})$. We assume an ordering of the jobs, which means that a decision $A_k$ can only depend on information available immediately prior to the decision, including the job's present characteristics $X_k$, and the history $H_{k-1}$ consisting of past job characteristics $X_{1:k-1}$, decisions $A_{1:k-1}$, and outcomes $Y_{1:k-1}$. This assumption is expressed more formally later.

\subsection{A Marked Point Process}
One of the foundations of causal inference is a causal model describing a probability model and causal relationships between variables. The variables are usually fixed and finite and either discrete or absolutely continuous. This poses a challenge for this paper's interest in a random number of jobs. To overcome this challenge, we propose using marked point processes (MPPs) as a probability model for the variables. This requires extending the definition of the random variables.

The first step in this extension is to attach a time $T_k$ to each job $k$, constrained so that $$0 < T_1 < T_2 \ldots < T_K < \infty,$$
where $T_k$ could be the observed or unobserved time at which the system first processes job $k$. To define times for all $k \in N$, we let $T_k = \infty$ whenever $k > K$. The $T_k$ can now comprise the \emph{times} of a marked point process (MPP), which must be an increasing (infinite) sequence taking values in $(0,\infty]$. The original random number $K$ of jobs can be recovered from the times $T_k$ by taking the supremum of the set $\{ k : T_k < \infty \}$.

The \emph{marks} of a marked point process also need to be defined for all $k \in \N$. This is done by taking $(X_k,A_k,Y_k)$ to be the marks of the process when $k \leq K$, and introducing an irrelevant mark $\Delta$ to extend the definition of these variables to all $k \in N$. Specifically, $X_k$, $A_k$, and $Y_k$ can take irrelevant marks $\Delta$ to signify that the corresponding time $T_k$ is infinite, or equivalently, that the panel size $K$ is less than $k$. The irrelevant mark is useful in subsequent derivations to remind us that it only makes sense to manipulate the marks numerically when $k \leq K$. Moreover, the original variables can be recovered perfectly from the $T_k$ and $(X_k,A_k,Y_k)$.

The second step of this extension is to define appropriate measurable spaces for the times and marks. For times up to a finite value $n$, we use the set
$${\cal X}^{(T_{1:n})} = \big\{ t_{1:n}\in \mathbb{R}^n : 0 < t_1 \leq \ldots \leq t_n;\, t_k < t_{k+1} \text{ if } t_k < \infty\big\},$$
which contains strictly increasing times while finite, and increasing times in general. Similarly, for an infinite sequence of times, we define the analogous set ${\cal X}^{(T_{1:\infty})}$.
For the marks, we use the sets
\begin{align*}
{\cal X}^{(X_{k})} = \mathbb{R}^l \cup {\Delta}, \quad
{\cal X}^{(A_{k})} = \{0,1\} \cup {\Delta}; \quad
{\cal X}^{(Y_{k})} = \mathbb{R} \cup {\Delta},
\end{align*}
which consist of marks that are either real-valued or the irrelevant mark $\Delta$. By taking products of these sets, additional sets such as ${\cal X}^{(X_{1:k})}$, ${\cal X}^{(X_{1:\infty})}$, ${\cal X}^{(H_k)}$, and ${\cal X}^{(H_{\infty})}$ can be defined. These spaces are equipped with appropriate Borel $\sigma$-algebras and product Borel $\sigma$-algebras, as described in \citet{jacobsen2006point}.

By extending the variable definition, we can model the distribution $\PR$ for $(T_k,X_k,A_k,Y_k)_{k \in \N}$. We can induce $\PR$ by constructing random variables iteratively from Markov kernels, which follows from the Ionescu Tulcea theorem (c.f. Theorem B.3.5. in \citet{hinderer2016dynamic} and Theorem~\sref{thm:ionescu_tulcea} in Appendix). Section~\ref{sec:causal_model} provides a more detailed specification of the Markov kernels. Once these kernels are properly defined, we can view the collection $(T_k, X_k, A_k, Y_k)_{k \in \N}$ as a \emph{marked point process}. This is a stochastic process that resides in 
${\cal X}^{(T_{1:\infty})} \times {\cal X}^{(H_{\infty})},$
for which the $T_k$ tend to infinity, and $X_k$, $A_k$, and $Y_k$ are the irrelevant mark if and only if $T_k$ is infinite.

\subsection{Causal Model} \label{sec:causal_model}

To place our random variables within a causal inference framework, we follow the approach of \citet{richardson2013single} and \citet{malinsky2019potential}, with one modification: we replace the use of conditional densities with Markov kernels. This allows us to handle an infinite collection of general random variables. A causal model is defined by specifying two objects for each random variable $Z$. The first object is a set of variables, denoted by $\Pa(Z)$, of \emph{parents} or \emph{direct causes} of $Z$. In general, the parents are those variables that, when intervened upon, directly affect $Z$. We make the following assumptions about the parents:

\begin{assumption}(Causal model - parents)\label{ass:parents}
For each $k  \in \N$, let
\begin{itemize}
    \item $\Pa(T_{k})= ( T_{1:k-1}, H_{k-1} )$
    \item $\Pa(X_{k})= ( T_{1:k}, H_{k-1} )$
    \item $\Pa(A_{k})=( X_{k}, H_{k-1} )$
    \item $\Pa(Y_{k})= ( X_{k}, A_{k},H_{k-1} ).$
\end{itemize}
These sets give rise to a directed acyclic graph (DAG), where the nodes represent variables and a directed edge exists from one variable to another if the former is a parent of the latter. 
\end{assumption}
To allow for greater generality, we tried not to impose any restrictions on which past observations could be direct causes of future observations. For instance, the parents of $T_k$ include all past observations: $T_{1:k-1}$ and $H_{k-1}$, and the same can be said for $X_k$. However, in order to identify the causal effects of interest, we had to impose some restrictions. For example, the parents of $Y_k$ and $A_k$ do not include the possibly unobserved times $T_{1:k}$.

\begin{figure}[ht!]
\centering
\caption{DAG induced by assumed causal model restricted to $(T_k,X_k,A_k,Y_k)_{k=1,2}$ when $K\geq 2$. \label{fig:dag}}
\begin{tikzpicture}[scale=0.75]
    \node[state] (t1) at (0.2,0) {$T_1$};
    \node[state] (x1) at (2.2,0) {$X_1$};
    \node[state] (a1) at (4.2,0) {$A_1$};
    \node[state] (y1) at (6.2,0) {$Y_1$};
    \path (t1) edge (x1);
    \path (x1) edge (a1);
    \path (a1) edge (y1);
    \node[state] (t2) at (0.2,-3) {$T_2$};
    \node[state] (x2) at (2.2,-3) {$X_2$};
    \node[state] (a2) at (4.2,-3) {$A_2$};
    \node[state] (y2) at (6.2,-3) {$Y_2$};
    \path (t2) edge (x2);
    \path (x2) edge (a2);
    \path (a2) edge (y2);
    \path (x1) edge (t2);
    \path (x1) edge (x2);
    \path (x1) edge (a2);
    \path (x1) edge (y2);
    \path (t1) edge (t2);
    \path (t1) edge (x2);
    \path (a1) edge (t2);
    \path (a1) edge (x2);
    \path (a1) edge (a2);
    \path (a1) edge (y2);
    \path (y1) edge (t2);
    \path (y1) edge (x2);
    \path (y1) edge (a2);
    \path (y1) edge (y2);
    \path (x1) edge[out=45,in=150,looseness=0.75] (y1);
    \path (x2) edge[out=-45,in=220,looseness=0.75] (y2);
\end{tikzpicture}
\end{figure}
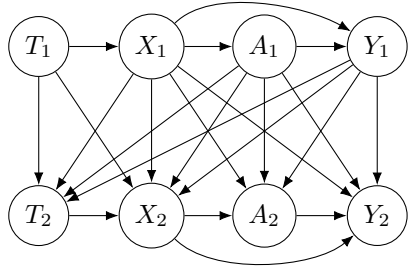

The DAG can be used to visualize the sets of parents. Although we cannot visualize the entire DAG for our variables, we can visualize the DAG restricted to a finite set. Figure~\ref{fig:dag} shows the DAG restricted to $(T_k,X_k,A_k,Y_k)_{k=1,2}$. It is worth noting that every DAG can be topologically ordered, which means that the nodes can be ordered in a way that ensures no node comes before its parents.

The second object is a Markov kernel from ${\cal X}^{(\Pa(Z))}$ to ${\cal X}^{(Z)}$ for each $Z$. These Markov kernels tell us how to construct a variable $Z$ in $(T_k,X_k,A_k,Y_k)_{k \in \N}$ from its parents. They also tell how to construct new variables were we to modify the value of its parents. To ensure we can easily move between each set of variables, it is desirable to define the same probability measure for both. Thus, in addition to assuming the existence of Markov kernels, it benefits us to regard the original and new variables as arising from the same collection of independent variables, often referred to as \emph{exogenous} variables. These considerations motivate defining Markov kernels as arising from a probability distribution for an exogenous variable and a deterministic assignment: 

\begin{assumption}(Causal model - Markov kernels)\label{ass:kernels} Assuming we have parents given by Assumption~\ref{ass:parents}, each variable $Z$ in the collection $(T_k,X_k,A_k,Y_k)_{k \in \N}$ is associated with a Markov kernel $\PR^{(Z)}$ from ${\cal X}^{(\Pa(Z))}$ to ${\cal X}^{(Z)}$. We assume this Markov kernel can be expressed as:
\begin{align*}
\PR^{(Z)}(B \, | \, \Pa_Z ) = \PR^{(\varepsilon_Z)} ( \{ w \in \Omega^{(Z)} \, | \, f^{(Z)}(\Pa_Z, w ) \in B \} ), 
\end{align*}
for some probability distribution $\PR^{(\varepsilon_Z)}$ 
on $(\Omega^{(Z)},{\cal E}^{(Z)})$ and a measurable function $f^{(Z)}$ from ${\cal X}^{(\Pa(Z))} \times \Omega^{(Z)}$ to ${\cal X}^{(Z)}.$ 
\end{assumption} 

To model $(T_k,X_k,A_k,Y_k)_{k \in \N}$ as an MPP, we require further conditions on the Markov kernels ensuring that, almost surely, the times are strictly increasing when finite and increasing otherwise, and the marks take the irrelevant mark exactly when the corresponding time is infinite. These assumptions are stated precisely in the Appendix (Assumption~\sref{ass:mpp_conditions}). Our causal model, now comprised of sets of parents and Markov kernels, gives us a probability model for the variables we are interested in:

\begin{assumption}(Causal model - distribution)\label{ass:probability_model}
Suppose that we have parents and kernels as given by Assumptions~\ref{ass:parents},\ref{ass:kernels}, and \sref{ass:mpp_conditions}. We obtain the distribution $\PR$ of a marked point process for $(T_k,X_k,A_k,Y_k)_{k \in \N}$ in two steps:
\begin{itemize}
\item Obtain a distribution $\PR$ on $(\varepsilon_{T_k},\varepsilon_{X_k},\varepsilon_{A_k},\varepsilon_{Y_k})_{k \in \N}$ using the Ionescu Tulcea theorem with respect to their respective sequence of probability distributions. 
\item Assign iteratively $Z=f^{(\varepsilon_Z)}(\Pa(Z), \varepsilon_Z)$ for each variable in ${(T_k,X_k,A_k,Y_k)}_{k \in \N}$
\end{itemize}
Finally, assume $K = \sup\{ k : T_k < \infty \}$ under $\PR$ is finite almost surely. 
\end{assumption}
It is important to mention that the ordering in which variables are assigned reflects a topological ordering of our variables that is compatible with our DAG. Specifically, each variable $Z$ must be defined after its parents are. In addition, we could have applied the Ionescu Tulcea theorem directly to the Markov kernels $\PR^{(Z)}$ to arrive at the same distribution $\PR$. However, as mentioned previously, it would be advantageous to fix the exogenous variables while manipulating the values of parents, so all our variables are defined with the same distribution $\PR$.

\subsection{Potential Outcomes}

Markov kernels describe the conditional distribution of a variable as a function of its parents. Crucially, by replacing the parents in the Markov kernel with fixed values, we can examine what would have happened under different scenarios. Specifically, we focus on examining what would have occurred if we had fixed the values of $A_{k}$:

\begin{definition}[Potential outcomes]\label{def:potential_outcomes}
Suppose we have $\PR$ on ${(\varepsilon_{T_k},\varepsilon_{X_k},\varepsilon_{A_k},\varepsilon_{Y_k})}_{k \in \N}$ from Assumption~\ref{ass:probability_model}. We define the potential outcomes for $B \subseteq \N$ and $a \in {\cal X}^{(A_{1:\infty})}$,
$$(T_k(a_B), X_k(a_B), A_k(a_B), Y_k(a_B))_{k \in \N},$$ 
iteratively according to $Z(a_B)=f^{(\varepsilon_Z)}(\Pa[a_B](Z), \varepsilon_Z),$ where we obtain $\Pa[a_B](Z)$ by replacing each variable $W$ in ${\Pa}(Z)$ either with $a_k$ if $W$ is $A_k$ and $k \in B$ or with $W(a_B)$ otherwise. The sets $\Pa[a_B](Z)$ give rise to a single world intervention graph (SWIG), where the nodes represent potential outcomes and the $a_k$, $k \in B$, and a directed edge exists from one variable to $Z(a_B)$ if the former is in the set $\Pa[a_B](Z)$. 
 \end{definition}

To put it another way, if the variable $A_k$ for $k \in B$ was originally used to construct a subsequent variable, the potential outcome definition dictates we pass forward $a_k$ instead of $A_k(a_B)$. Even though $A_k(a_B)$ does not get passed forward, we still construct $A_k(a_B),$ which distinguishes the causal inference framework of \citet{richardson2013single} from others. Moreover, while other frameworks may make certain assumptions about potential outcomes, assumptions like consistency end up being a natural consequence (c.f., \citet{malinsky2019potential}):

\begin{proposition}(Consistency)\label{prop:consistency}
Suppose we have potential outcomes defined for $B \cup C \subseteq \N$ and $C \subseteq \N$ with $B \cap C = \emptyset$ and $a \in {\cal X}^{(A_{1:\infty})}$. Among events for which $A_B(a_{B \cup C}) = a_B$, the potential outcomes $Z(a_B)$ and $Z(a_{B \cup C})$ are equal.
\end{proposition}

To clarify, consistency implies that potential outcomes $Z(a_{B \cup C})$ matches $Z(a_{C})$ when the potential outcome $A_B(a_{C})$ matches $a_B$. In essence, we need consistency so that $Y_k(a_B)$ has the same distribution as the original variable $Y_k$ when the intervention $A_B$ matches the value $a_B$ that we have fixed for the intervention.  This enables us to substitute $Y_k(a_B)$ for $Y_k$  when we condition on $A_B=a_B$. The proof of consistency follows by noting that, under the conditions of the proposition, $Z(a_B)$ and $Z(a_{B \cup C})$ are expressed identically in terms of the exogenous variables $\varepsilon^{(W)}$ and the assignment functions $f^{(W)}$ for $W$ in ${(T_k,X_k,A_k,Y_k)}_{k \in \N}.$ This identity allows us to claim the potential outcomes are equal \emph{everywhere}, as opposed to the weaker \emph{almost surely} or \emph{in distribution}.

The definition of potential outcomes leads us to another important concept called causal irrelevance, which tells us when we can drop the potential outcome notation. The SWIG induced by the potential outcome definition is the most convenient way to express causal irrelevance:

\begin{proposition}(Causal irrelevance)\label{prop:causal_irrelevance}
Suppose we have potential outcomes defined for $B \cup C \subseteq \N$ and $C \subseteq \N$ with $B \cap C = \emptyset$ and $a \in {\cal X}^{(A_{1:\infty})}$. If $Z(a_{B \cup C})$ is not a descendent of any $a_k$ with $k \in C$ in our SWIG, then the potential outcomes $Z(a_B)$ and $Z(a_{B \cup C})$ are equal.
\end{proposition}

In other words, if a potential outcome is not a descendant in the SWIG of an intervened value $a_k$, for $k \in B$, then the potential outcome is said to be causally irrelevant to the choice of $a_k$. In such cases, the value of the potential outcome does not change, whether or not we had passed $a_k$ forward at all. For example, since $T_1$ has no parents, then $T_1(a_B)$ is always equal to $T_1$. The proof of causal irrelevance is similar to that of consistency, i.e. under the conditions of the proposition, $Z(a_B)$ and $Z(a_{B \cup C})$ are expressed identically in terms of the exogenous variables and the assignment functions.

\begin{figure}[ht!]
\centering
\caption{SWIG restricted to $(T_k(a_1),X_k(a_1),A_k(a_1),Y_k(a_1))_{k=1,2}$ when $K\geq 2$.\label{fig:swig}}
\begin{tikzpicture}[scale=0.75]
\tikzset{elliptic state/.style={draw,ellipse}}
    \node[elliptic state] (t1) at (0.2,0) {$T_1$};
    \node[elliptic state] (x1) at (3.2,0) {$X_1$};
    \node[name=a1,shape=swig vsplit,swig vsplit={gap=3pt}] at (6.7,0) {    \nodepart{left}{$A_1$} \nodepart{right}{$a_1$} };
    \node[elliptic state] (y1) at (9.8,0) {$Y_1(a_1)$};
    \path (t1) edge (x1);
    \path (x1) edge (a1);
    \path (a1) edge (y1);
    \node[elliptic state] (t2) at (0.2,-3) {$T_2(a_1)$};
    \node[elliptic state] (x2) at (3.2,-3) {$X_2(a_1)$};
    \node[elliptic state] (a2) at (6.7,-3) {$A_2(a_1)$};
    \node[elliptic state] (y2) at (9.8,-3) {$Y_2(a_1)$};
    \path (t2) edge (x2);
    \path (x2) edge (a2);
    \path (a2) edge (y2);
    \path (x1) edge (t2);
    \path (x1) edge (x2);
    \path (x1) edge (a2);
    \path (x1) edge (y2);
    \path (t1) edge (t2);
    \path (t1) edge (x2);
    \path (a1) edge[out=-80,in=30,looseness=0.25] (t2);
    \path (a1) edge[out=-55,in=45,looseness=0.25] (x2);
    \path (a1) edge[out=-45,in=85,looseness=0.25] (a2);
    \path (a1) edge[out=-35,in=130,looseness=0.25] (y2);
    \path (y1) edge (t2);
    \path (y1) edge (x2);
    \path (y1) edge (a2);
    \path (y1) edge (y2);
    \path (x1) edge[out=30,in=155,looseness=0.75] (y1);
    \path (x2) edge[out=-30,in=205,looseness=0.75] (y2);
\end{tikzpicture}
\end{figure}
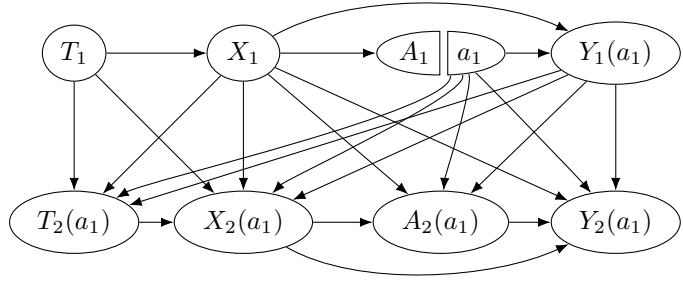

After applying causal irrelevance, we visualize the SWIG restricted to $(T_k(a_1),X_k(a_1),A_k(a_1),Y_k(a_1))_{k=1,2}$ with $B = \{1\}$ in Figure~\ref{fig:swig}. To obtain the SWIG from the original DAG, we split each node corresponding to $A_k$ for $k \in B$ into two nodes, one inheriting all incoming edges of $A_k$ and the other inheriting all outgoing edges. 

One of the key advantages of using SWIGs is that they uncover conditional independence relationships among potential outcomes. These relationships are a result of the SWIG being \emph{Markov compatible} with the potential outcome distribution, which then implies local and global Markov properties. The local Markov property, for example, says that a variable is independent of its non-descendants when conditioning on its parents.  The global Markov property characterizes conditional independence relationships arising from a property of the Markov compatible graph, known as \emph{d-separation}. While we will not delve into these concepts further in this paper, we will use a specific application of the global Markov property of the SWIG:

\begin{proposition}(Sequential ignorability)\label{prop:seq_ignorability}
Suppose we have potential outcomes defined for $B = \{k\}$ and $a \in {\cal X}^{(A_{1:\infty})}$. Then, $Y_{k+\ell}(a_k)$ is independent of $A_k$ conditional on $T_k(a_k) < \infty$ and any 
 vector $R_{k}(a_k)$ of variables that includes $H_{k-1}$, $X_k$, and a (possibly empty) subset of variables in $\Pa[a_k](Y_{k+\ell})$.
\end{proposition}

To demonstrate the above is an implementation of the global Markov property, we consider all paths in the SWIG that connect $A_k$ to $Y_{k + \ell}(a_k)$. Because $A_k$ has no outgoing edges (the SWIG shifts these edges to $a_k$), then each of these paths must go through the parents of $A_k$. These parents are exactly $X_k$ and $H_{k-1}$, which are included in $R_k$. In addition, it is important to note that the initial parent encountered by every path cannot be a collider. This is because $A_k$ is not considered a parent to any of its own parents. These conditions imply that $R_k(a_k)$ d-separates $A_k$ from $Y_{k + \ell}(a_k)$ in the SWIG, which in turn gives Proposition~\ref{prop:seq_ignorability}. We refer the reader to \citet{richardson2013single} and \citet{malinsky2019potential} for more on Markov properties of SWIGs.

\subsection{Lag Effect}

Now that we have our causal model in place and our potential outcomes defined, we can focus on what we want to learn from the data. Our aim is to understand the potential outcome of a future job if the current job is subjected to a specific intervention. Let $\ell > 0$ denote the time lag between the current job being intervened upon and a future job that might be affected by it. We refer to job $k+\ell$ as the \textit{future} job and job $k$ as the \textit{current} job. We can consider the potential outcomes as defined in Definition~\ref{def:potential_outcomes} with $B=\{k\}$ and $a_k \in \{0,1\}$ to capture this idea. According to causal irrelevance (Proposition~\ref{prop:causal_irrelevance}), we have $T_i(a_k)=T_i$, $X_i(a_k) = X_i$ and $A_i(a_k) = A_i$ when $i \leq k$; $Y_i(a_k)=Y_i$ when $i < k$.  Otherwise, we cannot drop the potential outcome notation. Provided it is not the irrelevant mark, we are interested in $Y_{k+\ell}(a_k)$, which represents the outcome for the future job we would have observed if the current job was forced to take intervention $a_k$. Although we cannot learn about individual potential outcomes $Y_{k+\ell}(a_k)$, we can estimate their expectations in various situations:

\begin{definition}[Lag effects]\label{def:bias_lag_effect}
Suppose we have potential outcomes given by Definition~\ref{def:potential_outcomes} for $B=\{k\}$ and $a_k = 1$ and separately for $a_k=0$. Let $R_{k}$ be any vector of variables that includes $H_{k-1}$, $X_k$, and any (possibly empty) subset of variables in $\Pa(Y_{k+\ell})\setminus A_k$. Introduce for $k, \ell \in \N$,
\begin{align*} 
\zeta(k,\ell,r) = \E\left[ Y_{k+\ell}(1)  | R_k(1) = r, T_{k+\ell}(1) < \infty \right] - 
    &\E\left[ Y_{k+\ell}(0)  | R_k(0) = r, T_{k+\ell}(0) < \infty \right]. 
\end{align*}
to represent the average causal effect on job $k+\ell$ after intervening on job $k$ among specific events for which $T_{k+\ell}(a_k) < \infty$ and $R_k(a_k)=r$ under the respective intervention.
\end{definition}

In the above definition, we compute expectation with respect to the probability distribution $\PR$, while holding the interventions of the first $k-1$ jobs at their original values and varying only the intervention for job $k$. The lag effect may depend on the index $k$ of the current job and the index of $k+\ell$ of the future job.  To avoid computing the irrelevant mark, we condition on $T_{k+\ell}(a_k)< \infty$. 

It is important to understand the causal query being asked with the lag effect, as it is subtle. We expose job $k$ to intervention $a_k=1$ and average the resulting value of $Y_{k+\ell}$ over all events which \emph{consequently} attain a certain level for $R_k$ (possibly post-intervention) and yield observations at time period $k+\ell$. This scenario is then contrasted to exposing job $k$ to intervention $a_k=0$ instead and averaging the resulting value of $Y_{k+\ell}$ over all events which consequently attain the same level for $R_k$ and yield observations at time period $k+\ell$. Our causal query is also distinctly different than what it would be were we to use:
\begin{align*}
\E\left[ Y_{k+\ell}(1)  | R_k = r, T_{k+\ell} < \infty \right] - 
    &\E\left[ Y_{k+\ell}(0)  | R_k = r, T_{k+\ell} < \infty \right].
\end{align*}
Above asks what would happen under different interventions among events who \emph{actually} attain a certain level of $R_k$ and yield observations at time period $k+\ell$ under the realized intervention. While neither are conventional, Pearl has a nice discussion on distinguishing between these two causal queries \citep{pearl2015conditioning} and why someone might condition on post-intervention variables. 

For us, we have chosen this lag effect based on practical considerations. One important consideration is that we need to condition on at least one post-intervention event, specifically the event where the induced $Y_{k+\ell}$ is observed. This is necessary to avoid computing expectations of irrelevant marks, and is unavoidable when the number of jobs, $K$, is random and influenced by interventions. This situation is common in many stochastic service systems, such as those that process jobs at different rates depending on prior interventions and in a fixed time frame (e.g., 9a to 5p).

Another factor to consider is that although the latter causal query, which conditions on realized values, may be more straightforward to comprehend, it is generally not identifiable. The latter query is equivalent to conditioning on colliders (e.g., $T_{k+\ell}$) between the intervention $A_k$ and the potential outcome $Y_{k+\ell}(a_k)$, thereby opening up a path to transmit non-causal associations from $A_k$ to $Y_{k+\ell}(a_k)$. The only way to close the path is to condition on potential outcomes related to post-intervention variables (such as $T_{k+\ell}(a_k)$). The two causal queries would coincide if the conditioning event $\{R_k(a_k) = r, T_{k+\ell}(a_k) < \infty\}$ remains invariant to $a_k$. For example, if $R_k$ only includes pre-intervention variables such as $H_{k-1}$ and $X_k$, and the intervention has no impact on whether the subsequent $Y_{k+\ell}$ is observed, then $\{R_k(a_k) = r, T_{k+\ell}(a_k) < \infty\}$ could be invariant to $a_k$. Yet, as noted above, we anticipate that the latter condition may be too restrictive for many stochastic systems.

A final consideration was what variables should we allow in the conditioning set. We choose to require $H_{k-1}$ and $X_k$ in $R_k$ for reasons that we will discuss in the next section, as they are relevant to identifying the causal effect. We allow for variables that occur post-intervention, but prior to $Y_{k+\ell}$, so we later can  better model the variation in $Y_{k+\ell}$. For example, the future job's characteristics $X_{k+\ell}$ may have a larger influence on the outcome $Y_{k+\ell}$ than the current job's characteristics $X_k$. We also consider these post-intervention variables to focus on the direct effects of the intervention, which fix the induced value of these variables to a certain level. For instance, we may wish to determine if directing a patient to vertical flow in the ED would speed up service for a future job, even were the induced congestion levels held constant.

Because the lag effect is high-dimensional, we may find it too difficult to specify the right functional form to express the influence of $r$ on the lag effect $\zeta(k,\ell,r)$. In these cases, it may be beneficial to use marginalization to reduce the dimension of the lag effect and then express the functional form for this lower-dimensional effect. One way to accomplish this is with the following:

\begin{definition}[Marginalized lag effects]\label{def:marginal_lag_effect}
Consider the lag effect $\zeta(k,\ell,r)$ as it is given in  Definition~\ref{def:bias_lag_effect}. Let $S_{k}$ be any (possibly empty) subset of $R_k$. Introduce
\begin{align*} 
\zeta_{\rm{marg}}(k,\ell,s) = \E\left[ \zeta(k,\ell,R_k) | S_k = s, T_{k+\ell} < \infty \right]. 
\end{align*}
to represent the lag effect marginalized over $R_k$ conditional on $S_k=s$ and $T_{k+\ell} < \infty$.
\end{definition}

If $S_k=R_k$, then the lag effect and the marginalized lag effect are equivalent. Our choice of the marginalized lag effect is also guided by practical considerations. In this case, our concern is not about identification. Identifying the marginalized lag effect can be achieved once the lag effect is identified. However, it is possible to condition on potential outcomes instead of observed variables. This would result in an identifiable effect, which can be calculated through a parametric g-formula \citep{robins1986new}. However, this approach requires a full specification of the Markov kernels, which we find too restrictive. As we saw before, if the conditioning event $\{R_k(a_k) = r, T_{k+\ell}(a_k) < \infty\}$ remains invariant to $a_k$, then the two strategies coincide, and the marginalized lag effect is simply:
\begin{align*}
\E\left[ Y_{k+\ell}(1)  | S_k = s, T_{k+\ell} < \infty \right] - 
    &\E\left[ Y_{k+\ell}(0)  | S_k = s, T_{k+\ell} < \infty \right].
\end{align*}

\subsection{Identification}\label{sec:bias_identification}

The primary challenge in causal inference is that we can only observe one of the potential outcomes, namely either $Y_{k+\ell}(1)$ or $Y_{k+\ell}(0)$. As a result, we are generally unable to estimate the lag effect (as defined in Definition~\ref{def:bias_lag_effect}). We need to extrapolate from our observations of $Y_{k+\ell}(1)$ when the intervention is assigned to $A_k=1$ in order to make inferences about $Y_{k+\ell}(1)$ when the intervention is assigned to $A_k=0$. For such extrapolation to be valid, we require the assumption of positivity in addition to consistency, causal irrelevance, and sequential ignorability (as outlined in Propositions~\ref{prop:consistency}--\ref{prop:seq_ignorability}). This assumption states that every job has a chance of being assigned to the intervention:

\begin{assumption}{(Positivity)}\label{ass:positivity} Consider $R_k$ as defined for the lag effect (Definition~\ref{def:bias_lag_effect}). Assume
$$\E[A_k|R_k,T_{k+\ell}<\infty]  \in (0,1)$$
almost surely.
    \end{assumption}

With positivity and the other properties, we are able to identify the lag effect, which is to say that we can re-express the lag effect equivalently in terms of observed variables:

\begin{lemma}{(Non-parametric identification)}\label{lem:identification} Consider the lag effect $\zeta(k,\ell,r)$ as it is given in  Definition~\ref{def:bias_lag_effect}. Assuming positivity (Assumption \ref{ass:positivity}), then the lag effect
$\zeta(k,\ell,r)$ is equivalent to
\begin{align*}
\E[Y_{k+\ell} \,|\, A_k=1, R_k=r, T_{k+\ell}< \infty] -
\E[Y_{k+\ell} \,|\, A_k=0, R_k=r, T_{k+\ell}< \infty].
\end{align*}
\end{lemma} 

\Pf
Sequential ignorability (Proposition \ref{prop:seq_ignorability}) and positivity (Assumption \ref{ass:positivity}) implies that
\begin{align*}
    \E[Y_{k+\ell}(a_k) \,|\, R_k(a_k)=r, T_{k+\ell}(a_k)< \infty] = \E[Y_{k+\ell}(a_k) \,|\, A_k=a_k, R_k(a_k)=r, T_{k+\ell}(a_k)< \infty].
\end{align*}
Consistency (Proposition \ref{prop:consistency}) simplifies the last expression to
\begin{align*}
     \E[Y_{k+\ell} \, | \, A_k=a_k, R_k=r, T_{k+\ell}<\infty].
\end{align*}
Applying these expressions to $a_k=1$ and $a_k=0$, we arrive at 
\begin{align*}
    \zeta(k,\ell,r) &= \E[Y_{k+\ell}(1) \,|\, R_k(1)=r, T_{k+\ell}(1)< \infty] - \E[Y_{k+\ell}(0) \,|\, R_k(0)=r, T_{k+\ell}(0)< \infty] \\
    &=\E[Y_{k+\ell} \,|\, A_k=1, R_k=r, T_{k+\ell}< \infty] - \E[Y_{k+\ell} \,|\, A_k=0, R_k=r, T_{k+\ell}< \infty],
\end{align*}
completing the proof. \EndPf

\section{Estimation}\label{sec:estimation}

\subsection{Estimating Equation}\label{sec:estimation_eqn}

Our goal is to estimate marginalized lag effects $\zeta_{\rm{marg}}(k,\ell,s)$ using estimating equations for a fixed $\ell \in \N$. Suppose we have collected data on $n$ panels. This data is assumed to be comprised of $n$ independent and identically distributed (iid) realizations of our marked point process $(T_k,X_k,A_k,Y_k)_{k \in \N}$ modeled with distribution $\PR$. We do not assume $T_k$ is observed or use irrelevant marks for estimation, so that observations are restricted to the set $(X_k,A_k,Y_k)_{k \in \{1,\ldots,K\}}$. Let $\E$ denote expectation with respect to $\PR$ and $\E_n$ denote the empirical average with respect to $n$ iid realizations of our marked point process.

The primary estimating function we work with is of the form
\begin{align} \label{eq:estimating_eqn}
\E_n[ U ]=0,    
\end{align}
where
\begin{align*}
    U =\sum_{k=1}^{K-\ell} U_k, \qquad \text{and} \qquad 
U_k = W_k \left( Y_{k+\ell} - g(R_k)'\alpha - A_k f(S_k)'\beta \right) \begin{bmatrix} g(R_k) \\ A_k f(S_k)\end{bmatrix}
\end{align*}
 for parameters $\alpha,\beta,\xi,\eta$ that we wish to estimate and suitable functions $g$ and $f$ that depend on $\xi$ and $\eta$. We search for parameters $\theta:=(\xi, \eta, \alpha,\beta)$ that solve Equation~\eqref{eq:estimating_eqn} in some compact subset $\Theta$ of $\mathbb{R}^m$. To simplify notation, we dropped the dependence of functions on parameters. The variable $U_k$ can be decomposed into several parts:
\begin{itemize}
    \item  $W_k$ (a weight for mitigating bias and emphasizing certain data points)
\item  $Y_{k+l}$ (outcome of future job)
    \item $g(R_k)'\alpha$ (a linear working model for the baseline conditional mean of $Y_{k+\ell}$)
    \item $f(S_k)'\beta$ (a linear working model for the marginalized lag effect)
    \item $\begin{bmatrix} g(R_k) \\ A_k f(S_k)\end{bmatrix}$ (derivative of the working mean model $g(R_k)'\alpha + A_k f(S_k)'\beta$ with respect to $\alpha$, $\beta$)
\end{itemize}
In particular, the parameter $\beta$ is the inferential target, as it contributes to the marginalized linear working model for the lag effect. If $K$ were deterministic, then the estimating equation (Equation~\ref{eq:estimating_eqn}) has several precedents. It is akin to a generalized estimating equation (GEE), in which data are weighted, clustered into panels, and given an independent working correlation matrix. It is also akin to the g-estimation approach of \citet{vansteelandt2016revisiting} for structural nested mean models and the weighted and centered approach of \citet{boruvka_assessing_2018}.

While not the inferential target, the estimating equation depends on parameters $\xi$ and $\eta$. These parameters specify a working model for conditional probabilities of job $k$'s intervention assignment:
\begin{align*}
 q_k(S_k; \xi) &\approx \E[ A_k \, | \, S_{k}, T_{k+\ell} < \infty] \\
p_k(R_k; \eta) &\approx \E[ A_k \, | \, R_k, T_{k+\ell} < \infty]
\end{align*}
We assume $\xi$ is also estimated using an estimating equation of the form
\begin{align*} 
\E_n[ L ]= \E_n\left[ \sum_{k=1}^{K-\ell} L_k \right] = 0,  
\end{align*}
for variables $L_k \mathbbm{1}_{\{k \leq K\}}$ that depends only on $\xi$ and the observations of $(X_k,A_k,Y_k)_{k \in \{1,\ldots,K\}}$. Similarly, we assume $\eta$ is also estimated using an estimating equation of the form
\begin{align*}
\E_n[ M ]= \E_n\left[ \sum_{k=1}^{K-\ell} M_k \right] = 0
\end{align*}
for variables $M_k\mathbbm{1}_{\{k \leq K\}}$ that depends only on $\eta$ and the observations of $(X_k,A_k,Y_k)_{k \in \{1,\ldots,K\}}$. Most likely, $\xi$ and $\eta$ would be estimated using logistic or log-linear regression. For example, if we used logistic regression and assumed a linear model in $S_{k}$ on the log-odds scale with $\xi$ as the regression coefficients, our estimating equation might use:
$$ L = \sum_{k=1}^{K-\ell} (A_{k} - \mathrm{logit}^{-1}([1,S_{k}]'\xi)) \begin{bmatrix} 1 \\ S_k \end{bmatrix}. $$

By stacking the estimating equations together,
\begin{align} \label{eq:stacked_eqn}
    \E_n\left\{ \begin{bmatrix} U \\ L \\ M \end{bmatrix} \right\} = \E_n\left\{ 
    \sum_{k=1}^{K-\ell}\begin{bmatrix} U_k \\ L_k \\ M_k \end{bmatrix} \right\} = 0,
\end{align}
estimation can be viewed as searching for $\alpha$, $\beta$, $\xi$, and $\eta$ that solve the stacked estimating equation. If we define $U_{\rm{stacked}}$ as the concatenated vector of $U$, $L$, and $M$, then in order for Equation~\eqref{eq:stacked_eqn} to be meaningful, we are assuming that $U_{\rm{stacked}}$ is integrable. Since these variables are sums of a random number of variables, we impose mild conditions on $K$ and $U_k$, $L_k$, and $M_k$ to recover integrability (see Assumption~\sref{ass:int_regularity} in Appendix) to recover integrability:

\begin{lemma}[Integrability]\label{lem:integrability} Under the conditions on $U_k$, $L_k$, $M_k$ and $K$ in Assumption~\sref{ass:int_regularity}, Equation~\ref{eq:stacked_eqn} is well-defined in the sense that, for each $\theta \in \Theta$, $U_{\rm{stacked}}$ is integrable and has finite expectation. 
\end{lemma}
The assumptions can be found in Appendix \sref{app:regularity}, and the proof in Appendix \sref{app:integrability}. The broad arc is to replace the random sum in the definition of $U_{\rm{stacked}}$ with an infinite sum and then exchange the expectation and the infinite sum. The conditions in Assumption~\sref{ass:int_regularity} are then used to ensure that the final sum can be bounded above by the expectation of $K$ up to a constant.

\subsection{Estimating Procedure}\label{sec:est_procedure}

The first step of estimation would be to use the observations of $(X_k,A_k,Y_k)_{k \in \{1,\ldots,K\}}$ to recover an estimate $\xi_n$ and $\eta_n$ that solves the estimating equations $\E_n[U]=\E_n[V]=0$. For example, we might use logistic regression to model the above conditional probabilities on the log odds scale as a linear function of the $S_k$ or $R_k$. In such a case, the parameters $\xi_n$ and $\eta_n$ would be the estimated regression coefficients. In special circumstances, the intervention $A_k$ was randomized according to a known probability.  Consequently, $q_k(S_k; \xi)$ and $p_k(R_k; \eta)$ would be exact. 

Once $\xi$ and $\eta$ are estimated, we turn to specifying $W_k$, $f$, and $g$. The weight term $W_k$ is taken to be the ratio of the two conditional probability models:
\begin{align}\label{eq:weig}
    W_k = A_k \frac{ q_k(S_k; \xi_n) }{ p_k(R_k; \eta_n) } + (1-A_k)\frac{ 1-q_k(S_k; \xi_n) }{1-p_k(R_k; \eta_n) }.
\end{align}
Naturally, we are assuming that we are not dividing by zero, otherwise the weights would be ill-defined. This assumption is the empirical equivalent of our positivity assumption (Assumption~\ref{ass:positivity}). These weights resemble (stabilized) inverse probability weights, in which a job is weighted inversely according to the probability of receiving their realized intervention assignment \citep{horvitz1952generalization}. Jobs that are more likely to receive their realized intervention assignment are subsequently down-weighted relative to those that are less likely. Given that certain jobs may be predisposed to be assigned a particular intervention, the weights create a pseudo-population of jobs in each intervention group that better reflects all jobs, not just those in the given intervention group. In the special case when $S_k = R_k$ or when $p_k(R_k;\eta_n)=q_k(S_k; \xi_n)$, then the weights $W_k$ are just $1$. 

Meanwhile, we specify $f(S_k)$ to be any vector-valued function of $S_k$. It reflects our assumptions about how we think the marginalized lag effect might vary as a function of $S_k$, i.e.
\begin{align*}
    f(S_k)'\beta \approx \xi_{\rm{marg}}(k,\ell,S_k) 
\end{align*}
For example, we might think the marginalized lag effect of assigning a patient to vertical flow is quadratic in a patient's age. If $S_k$ were the current patient's age, then we might chose:
\begin{align*}
f(S_k) = \begin{bmatrix} 1 & S_k & S_k^2 \end{bmatrix}'
\end{align*}
We can similarly specify the baseline term $g(R_k)$ to be any vector-valued function of $R_k$ with one caveat. It must contain the variable $q_k(S_k; \xi) f(S_k)$. This caveat is so that the working mean model:
\begin{align*}
    g(R_k)'\alpha + A_k f(S_k)' \beta,
\end{align*}
upon a re-definition of $\alpha$, includes the model in which the $A_k$ is centered:
\begin{align*}
    g(R_k)'\alpha + (A_k - q_k(S_k;\xi_n)) f(S_k)' \beta
\end{align*}
Including $q_k(S_k; \xi) f(S_k)$ in $g(R_k)$ follows the strategy of \citet{vansteelandt2016revisiting} of making the mean model more general while still, practically, centering the intervention assignments. Centering is recommended for reasons we discuss later and is recommended in \citet{boruvka_assessing_2018}.

The last step of estimation is to search for a solution $\alpha_n$ and $\beta_n$ to the estimating equation (Equation~\ref{eq:estimating_eqn}). Given the similarity of Equation~\eqref{eq:estimating_eqn} to GEE, these parameters (though not necessarily their standard errors) can be estimated using standard GEE software, provided the working correlation matrix is specified to be independent.  We next study the properties of the resulting estimator $\beta_n$.

\subsection{Consistency and Asymptotic Normality}\label{sec:consistency}

Our estimator is a type of estimator called a Z-estimator (Z for ``zero"), as it involves searching for roots of an estimating equation. Asymptotic properties of Z-estimators have been extensively studied, so that we can invoke standard arguments. In an effort to be self-contained, we sketch the arguments in \citet{van2000asymptotic}. We start with the following definition:

\begin{definition}[Estimator] \label{def:estimator}
Define an estimator $\theta_n = (\xi_n,\eta_n,\alpha_n,\beta_n)$ to be a solution in a set $\Theta \subseteq \mathbb{R}^m$ to the (stacked) estimating equation (Equation~\ref{eq:stacked_eqn}), should such a solution exist. Define $\theta_{\infty}=(\xi_{\infty},\eta_{\infty},\alpha_{\infty},\beta_{\infty})$ to be a solution in $\Theta$ to the asymptotic version of the estimating equation:
$$ \E\left[ U_{\rm{stacked}}\right] = 0,$$
should such a solution exist.
\end{definition}

The first property of our estimators is consistency:

\begin{theorem}[Consistency] \label{thm:asymp_consistency}
Under Assumption~\sref{ass:cons_regularity},  the estimator $\theta_n$ in Definition~\ref{def:estimator} is consistent with respect to $\theta_{\infty}$:
$$ \theta_n \stackrel{P}{\rightarrow} \theta_{\infty}.$$
\end{theorem}

\Pf
The proof follows from Theorem 5.7 in \citet{van2000asymptotic} if two conditions are met. Take $M_n(\theta) = -\begin{Vmatrix} \E_n [U_{\rm{stacked}}] \end{Vmatrix}$ and $M(\theta) = -\begin{Vmatrix} \E[U_{\rm{stacked}}] \end{Vmatrix}$. The first condition is the uniform convergence of $M_n(\theta)$ to $M(\theta)$ in probability over $\Theta$. Given that $\E_n [U_{\rm{stacked}}]$ is the empirical average over iid samples of $U_{\rm{stacked}}$, this uniform convergence is an application of a uniform law of large numbers and follows from several conditions, including compactness of $\Theta$ in $\mathbb{R}^m$, finiteness of $\E[U_{\rm{stacked}}]$ (Lemma~\ref{lem:integrability}), and the mapping $\theta \rightarrow U_{\rm{stacked}}$ being almost surely continuous and dominated by an appropriate function (see Lemma 1 in \cite{tauchen1985diagnostic}). The second condition is called well-separatedness and requires that the supremum of $M(\theta)$ over all $\Theta$ outside of any neighborhood around a unique maximizer $\theta_{\infty}$ is positive. Well-separatedness follows immediately from the uniqueness of the maximizer and the almost sure continuity of the mapping $\theta \rightarrow U_{\rm{stacked}}$ over compact $\Theta$  (Assumptions~\sref{ass:cons_regularity}). The proof of Theorem 5.7 in \citet{van2000asymptotic} uses the uniform convergence condition to argue that $M(\theta_n) \stackrel{P}{\rightarrow } M(\theta_{\infty})=0$. Well-separatedness then ensures that the only way $M(\theta_n)$ converges in probability to $0$ would be if $\theta_n$ were converging in probability to $\theta$. 
\EndPf

The assumptions required for consistency (Assumption~\sref{ass:cons_regularity}) in common choices of $L$ and $M$ (such as logistic or log-linear models) are relatively mild. Continuity of $\theta \rightarrow U_{\rm{stacked}}$ is reasonable, since $U$ is linear in $\alpha$ and $\beta$ and smooth in $\xi$, and $L$ and $M$ is constant in $\alpha$ and $\beta$ and smooth in $\xi$ and $\eta$. The main concern is that $U$ involves division by $p_k(R_k;\eta)$ or $1-p_k(R_k;\eta)$, which could prevent continuity over $\Theta$ if $p_k(R_k;\eta)$ is not bounded away from $0$ and $1$. For a function to dominate the mapping $\theta \rightarrow U_{\rm{stacked}}$, $K$ and variables contributing to $U_k$ cannot be too large. This is comparable to the conditions for integrability (Assumption~\sref{ass:int_regularity}). In practice, these variables are usually bounded. Similarly, parameters are usually bounded in practice, so compactness of $\Theta$ is usually not an issue.

The second property of our estimators is $\sqrt{n}$-consistency and asymptotic normality:

\begin{theorem}[Asymptotic normality] \label{thm:asym_normality}
Under Assumption~\sref{ass:amvn_regularity}, the estimator $\theta_n$ in Definition~\ref{def:estimator} is $\sqrt{n}$-consistent with respect to $\theta_{\infty}$:
$$ \sqrt{n}(\theta_n -\theta_{\infty}) = - \mathbb{B}^{-1} \sqrt{n} E_n\left[ U_{\rm{stacked}} \big|_ {\theta = \theta_{\infty}}\right] + o_p(1)$$
and asymptotically normal with mean $\theta_{\infty}$ and variance $\mathbb{B}^{-1} \mathbb{C} (\mathbb{B}^{-1})'$, where
\begin{align*}
\mathbb{B} = \E\left[ \nabla_{\theta}U_{\rm{stacked}} \big|_ {\theta = \theta_{\infty}}\right]; \qquad 
\mathbb{C} = \E\left[U_{\rm{stacked}}U_{\rm{stacked}}' \big|_ {\theta = \theta_{\infty}} \right].
\end{align*}
\end{theorem}

\Pf
The proof follows from several results in \citet{van2000asymptotic}, including Theorem 5.21, Lemma 19.24, and Example 19.7. Let $\Psi(\theta) = \E[ U_{\rm{stacked}} ]$ and $\Psi_n(\theta) = \E_n[ U_{\rm{stacked}}]$, such that $\Psi(\theta_{\infty})=\Psi_n(\theta_n) = 0$. We can obtain an approximation for $\sqrt{n} \Psi(\theta_n)$ from $\Psi(\theta_{\infty})=0$ and the differentiability of $\Psi(\theta)$ at $\theta_{\infty}$:
\begin{align*}
\sqrt{n} \Psi(\theta_n) = \sqrt{n} \, \mathbb{B} \cdot (\theta_{n} - \theta_{\infty}) + \sqrt{n} \begin{Vmatrix} \theta_n - \theta_{\infty} \end{Vmatrix} o_P\left(1\right).
\end{align*}
Meanwhile, the central limit theorem allows us to conclude that $\sqrt{n} \Psi_n(\theta_{\infty})$ is asymptotically normal with mean zero and covariance $\mathbb{C}.$ The challenge is to connect the two aforementioned results through empirical process theory, which necessitates demonstrating that $\sqrt{n} \Psi_n(\theta_{\infty}) + \sqrt{n} \Psi(\theta_n) \stackrel{P}{\rightarrow} 0$. This connection relies on the compactness of $\Theta$ and the (almost sure) Lipschitz continuity of $\theta \rightarrow U_{\rm{stacked}}$, which provides uniform convergence in distribution over $\Theta$ of the empirical process $\sqrt{n}\left(\Psi_n(\theta) - \Psi(\theta)\right)$. Thus, we can deduce:
\begin{align*}
\sqrt{n} \, \Psi(\theta_n) = \sqrt{n} \, \mathbb{B} \cdot (\theta_{n} - \theta_{\infty}) + \sqrt{n} \, \begin{Vmatrix} \theta_n - \theta_{\infty} \end{Vmatrix}\,o_P\left(1\right) = -\sqrt{n} \, \Psi_n(\theta_{\infty}) + o_P(1).
\end{align*}
Last, we argue $\sqrt{n} \, \begin{Vmatrix} \theta_n - \theta_{\infty} \end{Vmatrix}\,o_P\left(1\right) = o_P(1)$ because $\sqrt{n} \Psi_n(\theta_{\infty}) = O_P(1)$ and $\mathbb{B}$ is invertible.
\EndPf

The additional assumptions required for asymptotic normality (Assumption~\sref{ass:amvn_regularity}) in common choices of $L$ and $M$ (such as logistic or log-linear models) are relatively mild. Similar to continuity, it is reasonable to assume Lipschitz continuity and differentiability for the mapping $\theta \rightarrow U_{\rm{stacked}}$, since $U_{\rm{stacked}}$ is a smooth function over $\Theta$ with $p_k(R_k;\eta)$ bounded away from $0$ and $1$ if needed. Finiteness of $\mathbb{B}$ is linked to integrability and finding a dominating function for $\theta \rightarrow U_{\rm{stacked}}$, and requires that $K$ and the variables contributing to $U_k$ cannot be too large. Invertibility of $\mathbb{B}$ follows from the uniqueness of our minimizer $\theta_{\infty}$. The main concern is the finiteness of $\mathbb{C}$ and the square-integrability of the Lipschitz constant, which further limits the size of $K$ and the variables contributing to $U_k$.

\subsection{Double Robustness}\label{sec:double_robust}

Our next task is to connect $\beta_{\infty}$, for which $\beta_n$ is consistent, to the marginalized lag effect $\xi_{\rm{marg}}(k,\ell,s)$. We can make this connection if some of our various modeling assumptions is correct.  All told, there are four different models that we wish were valid for some $\theta \in \Theta$: 
\begin{itemize}
    \item Outcome model, viz,
\begin{align}
    g(R_k)'\alpha + A_k f(S_k)'\beta = \E[Y_{k+\ell} \,| \, R_k, A_k, T_{k+\ell} < \infty]  \label{eq:outcome_model},
\end{align}
    
\item Conditional probability model in the numerator of the weights, viz,    
    \begin{align}
    q_k(S_k;\xi) = \E[A_{k} \,| \, S_k, T_{k+\ell} < \infty], \label{eq:numerator_model}
    \end{align}
    \item Conditional probability model in the denominator of the weights, viz,    
    \begin{align}
    p_k(R_k;\eta) = \E[A_{k} \,| \, R_k, T_{k+\ell} < \infty], \label{eq:denominator_model}
    \end{align}
    
     \item Marginalized lag effect, viz,    
    \begin{align}
    f(S_k)'\beta = \zeta_{\rm{marg}}(k,\ell,S_k), \label{eq:lag_model}
    \end{align}    
\end{itemize}

\begin{theorem}[Double robustness] \label{thm:double_robust}
Suppose positivity (Assumption~\ref{ass:positivity}) and the conditions for asymptotic normality (Theorem~\ref{thm:asym_normality}) hold. If either the outcome model is correct (Equation~\ref{eq:outcome_model}) or the conditional probability model in the denominator of weights (Equation~\ref{eq:denominator_model})
is correct for $\theta_{\infty} \in \Theta$, then the asymptotic limit of our estimator $\beta_{\infty}$ is related to the lag effect $\zeta_{\rm{marg}}(k,\ell,S_k)$ according to the linear equation:
$$\mathbb{G}\beta_{\infty} = g$$
where 
\begin{align*}
\mathbb{G} &= \sum_{k=1}^{\infty} \E\left[ q(S_k; \xi_{\infty})(1-q(S_k; \xi_{\infty}))f(S_k)f(S_k)' \,\big|\, T_{k+\ell} < \infty\right] \PR( T_{k+\ell} < \infty )  \\
g &= \sum_{k=1}^{\infty} \E\left[ q(S_k; \xi_{\infty})(1-q(S_k; \xi_{\infty}))\zeta_{\rm{marg}}(k,\ell,S_k) f(S_k) | T_{k+\ell} < \infty \right] \PR( T_{k+\ell} < \infty ).
\end{align*}
If, in addition, $\mathbb{G}$ is invertible and the lag effect model is correct (Equation~\ref{eq:lag_model}) for some $\theta_* = (\xi_*,\eta_*,\alpha_*,\beta_*) \in \Theta$, then 
$$\beta_{\infty} = \beta_*.$$
\end{theorem}

The proof for this theorem can be found in Appendix \sref{app:double_robust}. There are several points to consider. First, $\beta_{\infty}$ can be viewed as a weighted average of the causal effect $\zeta_{\rm{marg}}(k,\ell,S_k)$, even if the lag effect model is incorrect. This implies that we have a third form of robustness where, although we may not precisely recover our target effect, we can still obtain a relevant causal effect. We use a weighting scheme based on $q_k(S_k; \xi_{\infty})(1-q_k(S_k; \xi_{\infty}))$. This idea is not novel, as there exists a recent method in causal inference known as overlap weights. This method weights the average causal effect conditional on a variable $x$, denoted by $\mu(x)$, by $e(x)(1-e(x))$, where $e(x)$ is the propensity score conditional on $x$ \citep{li2019addressing}. Our weighting scheme gives greater importance to jobs where $S_k$ is balanced between intervention groups. Such jobs have intervention decisions that are essentially random, allowing for greater comparability between the groups. Last, double robustness is hiding an important point. The error terms in the proof are directly related to the errors in both the conditional probability model in the denominator (Equation~\ref{eq:denominator_model}) and the outcome model (Equation~\ref{eq:outcome_model}). Even if both models are incorrect but are good approximations, $\mathbb{G}\beta_{\infty} \approx g.$

\subsection{Semiparametric Efficiency}\label{sec:efficiency}

The theorem on asymptotic normality (Theorem~\ref{thm:asym_normality}) provides some notion of efficiency. It shows that the estimators' variance takes the form of the standard sandwich estimator $\frac{1}{n}\mathbb{B}^{-1} \mathbb{C} (\mathbb{B}^{-1})'$. This observation enables us to estimate the variance of our estimator, along with standard errors, by using their empirical counterpart:
\begin{align*}
\mathbb{B}_n = \E_n\left[ \nabla_{\theta}U_{\rm{stacked}} \big|_ {\theta = \theta_{n}}\right]; \qquad 
\mathbb{C}_n = \E_n\left[U_{\rm{stacked}}U_{\rm{stacked}}' \big|_ {\theta = \theta_{n}} \right].
\end{align*}

One important question is whether there is a better way to estimate the inferential target $\beta$. This question is often asked in terms of asymptotic efficiency, which captures the extent to which our estimator for $\beta$ has small asymptotic variance. 
To compute the asymptotic variance of our estimator of $\beta$, we can use the matrix $\frac{1}{n}\mathbb{B}^{-1} \mathbb{C} (\mathbb{B}^{-1})'$. In particular, the asymptotic variance of our estimator for $\beta$ is given by the $\beta$ portion of the formula, which can be written as:
$$\frac{1}{n} \left( \nabla_{\beta} \theta \cdot \mathbb{B}^{-1} \right) \mathbb{C} \left( \nabla_{\beta} \theta \cdot \mathbb{B}^{-1} \right)'.$$
Although our estimator is not generally optimal, we offer the following evidence to suggest that it is a sensible choice:

\begin{theorem}[Efficient score] \label{thm:opt_efficiency}
Consider the special case when $S_k=R_k=\Pa(Y_{k+\ell})\setminus A_k$ and the semi-parametric model ${\cal M}$ described in Definition~\sref{def:modeling_space}. Let $\E_0$ and $\rm{Var}_0$ denote expectation and variance with respect to the correct model in ${\cal M}.$ Then, the efficient score function for $\beta$ with respect to ${\cal M}$ is 
\begin{align*}
    S_{\rm{eff}}(\beta) = \sum_{k=1}^{K-\ell} \frac{f(R_k)}{\sigma_k(R_k)}\left( Y_{k+\ell} - \mu_k(R_k) - A_k f(R_k)'\beta \right) \left( A_k - \rho_k(R_k) \right)
\end{align*}
where 
\begin{equation*}
\begin{aligned}
\rho_k(R_k)   &= \E_0\left[A_k | R_k, T_{k+\ell} < \infty \right] \\
\mu_k(R_k) &= \E_0[Y_{k+\ell} | A_k=0, R_k, T_{k+\ell} < \infty ] \\
    \sigma_k(R_k) &= (1-\rho_k(R_k)) {\rm{Var}}_0[Y_{k+\ell} | A_k=1,R_k,T_{k+\ell}<\infty] + \rho_k(R_k) {\rm{Var}}_0[Y_{k+\ell} | A_k=0,R_k,T_{k+\ell}<\infty].
\end{aligned}
\end{equation*}

\end{theorem}

The proof of the theorem above is presented in Appendix \sref{app:efficiency}. It is worth noting that the condition $S_k = R_k$ implies $W_k=1$, based on Equation \ref{eq:weig}. The form of the efficient score inspired us to define the estimating equation (Equation~\ref{eq:estimating_eqn}). Specifically, under the conditions stated in the theorem, our proposed estimator satisfies the following expression:
$$\E_n\left[ \sum_{k=1}^{K-\ell} f(R_k)\left( Y_{k+\ell} - g(R_k)'\alpha - A_k f(R_k)'\beta \right) \left( A_k - p_k(R_k;\eta) \right) \right].$$
Upon comparing this expression to the one in the theorem, we observe that although we do not have precise knowledge of $\mu_k(R_k)$, we utilize $g(R_k)'\alpha$ as an approximation for $\mu_k(R_k)$. Similarly, while we may not have exact information about $\rho_k(R_k)$, we use $p_k(R_k;\eta)$ as an approximation for $\rho_k(R_k)$. These ideas are then used for the general case when the conditioning set $S_k$ is a proper subset of $R_k$ or when $R_k$ is a proper subset of $\Pa(Y_{k+\ell})\setminus A_k.$

\section{Case Study}\label{sec:bias_case_study}

We apply our developed framework to observational data obtained from a large academic hospital in the Midwest, where a split-flow model has been implemented in the ED. In this increasingly adopted flow model, nurses at triage are replaced by physicians who assess whether patients should be directed to a vertical area for low-acuity cases or a traditional room. Existing empirical evidence primarily focuses on evaluating the overall effects of split-flow on ED operations and patient outcomes~\citep{konrad2013modeling, wiler2016implementation, garrett2018effect}. However, the specific impact of the decision to assign a patient to the vertical area on subsequent patient outcomes remains uncertain.

To estimate lag effects, we use data on $n=27831$ encounters with an ED split flow model between November 1, 2016, and September 26, 2018. Since split-flow is operated within a specific time window each day, we treated encounters on a given day as independent panels. We excluded the 5 days that had fewer than 3 ED split-flow visits. Characteristics $X_k$ of patient $k$ within a panel consisted of
\begin{itemize}
\item Demographic information on age, gender, and race,
\item ED census, representing the number of patients in the ED upon arrival, 
\item Chief complaint, categorized into four common complaints: abdominal pain, chest pain, dyspnea, and headache. There was an additional category to capture the remaining complaints;
\item Comorbidity factors including the Hierarchical Condition Category (HCC) score~\citep{RTC-Dec2018} and binary indicators of congestive heart failure, hypertension, obesity, diabetes with and without complications, and hypertension. 
\end{itemize}
The intervention $A_k$ of patient $k$ is whether the patient is assigned to the vertical area ($A_k=1$) or a traditional bed ($A_k=0$). We considered four outcomes $Y_k$, with the estimation procedure repeated for each outcome. These outcomes included time to disposition after being roomed (in minutes), number of tests performed such as electrocardiograms and radiology scans, admission decision, and the patient routing decision to either vertical area or fast track area (which is simply $A_k$). 

The lag $\ell$ was fixed at 1. We defined $R_k$ to include characteristics $X_{k+\ell}$ of the future patient. To model the conditional probability $p_k(R_k;\eta)$ in the denominator of the weights, we performed GEE regression with a logit link function and an independent working correlation matrix. The regression model included the outcome $A_{k}$ as the dependent variable. Independent variables included the current patient's characteristics $X_k$, the future patient's characteristics $X_{k+\ell}$, and a fixed number $i$ of lagged decisions $A_{k-i},\ldots,A_{k-1}$. Quadratic and cubic splines were considered for age of the current patient with knots placed at tertiles in the age distribution (i.e., 30, 46, and 62 years). Model comparison was conducted using the Quasi-likelihood under the Independence model Criterion (QICu)~\citep{pan2001akaike}, where different choices in $i$ and age splines were evaluated. The model with the smallest QICu value was selected, which consisted of $i=4$ and quadratic splines for age.

We explored various choices for $S_k$. We began with $S_k$ being an empty set ($\emptyset$). To demonstrate the flexibility of considering different $S_k$, we then set $S_k$ to include each of the comorbidity variables for the future patient, which are available in $X_{k+\ell}$. To model the conditional probability $q_k(S_k;\xi)$ in the numerator of the weights, we performed GEE regression with a logit link function and an independent working correlation matrix. In the regression model, we treated $A_{k}$ as the dependent variable. When $S_k$ was not empty, it was included as an independent variable in the model.

For each outcome and each $S_k$, we estimated the inferential target of $\beta$. In the context of Theorem~\ref{thm:double_robust}, $f(S_k)'\beta$ corresponds to the marginalized lag effect $\zeta_{\rm{marg}}(k,\ell,S_k)$. To achieve this, we solved the estimating equation presented in Equation \eqref{eq:estimating_eqn} for a given baseline model $g(R_k)'\alpha$ and marginalized lag effect model $f(S_k)'\beta$. The term $f(S_k)$ included a constant term of $1$ to represent the main intervention effect, and when $S_k$ was one of the comorbidity variables, it incorporated $S_k$ as well. Our baseline model included a constant term of $1$ to represent the intercept, as well as the characteristics of the current ($X_k$) and future ($X_{k+\ell}$) patient. Additionally, the requisite term $f(S_k)q_k(S_k;\xi_n)$ was included, where $\xi_n$ represents the estimate obtained from performing GEE regression to model $q_k(S_k;\xi)$. As mentioned earlier, we tackled the estimating equation by employing an equivalent procedure of GEE regression with an identity link and independent working correlation matrix. 

Tables~\ref{tab:results_operational} and \ref{tab:results_binary} reports the main effect $\beta$ when $S_k=\emptyset$ and the interaction term $f(1)'\beta-f(0)'\beta$ when $S_k \neq \emptyset$. Standard errors were calculated using the \textit{geex} package \citep{geex}. The implementation details are in Appendix \sref{app:geex}. Estimates and standard errors were used to construct Wald 95\% confidence intervals and perform Wald hypothesis tests of the null hypothesis that the true value of the estimate is zero. Significance was considered $P < .05$. Given the exploratory nature of this investigation, marginal significance was considered $P < .10$.

\begin{table}[ht]
    \footnotesize
    \centering
    \caption{Estimates for main effect ($S_k = \emptyset$) or interaction term ($S_k \neq \emptyset$) of sending the current patient to the vertical area on the time to disposition and number of tests for the next patient. We report $P$ values for Wald tests.\label{tab:results_operational}}
    \begin{tabular}{l l l l l }
        \toprule
        & \multicolumn{2}{c}{\textbf{Time to disposition (min.) }} & \multicolumn{2}{c}{\textbf{Number of tests}} \\
        \cmidrule(lr){2-3} \cmidrule(lr){4-5} 
         \textbf{Variable $S_k$} & Estimate (95\% CI) & $P$ & Estimate (95\% CI) & $P$ \\
        \midrule
        $\emptyset$ & ~~3.6 (-1.1, 8.3) & .11 & ~0.04 (0.01, 0.09) & .02 \\
        Congestive heart failure & ~-3.1 (-22.3, 16.1) & .74 & -0.06 (-0.2, 0.1) & .56 \\
        Hypertension & ~-1.9 (-11.0, 7.2) & .68 & -0.03 (-0.1, 0.05) & .43 \\
        Obesity & ~~2.9 (-9.4, 15.4) & .63 & ~0.02 (-0.1, 0.2) & .70\\
        Diabetes with complications & -10.5 (-25.4, 4.4) & .16 & -0.01 (-0.1, 0.1) & .88\\
        Diabetes without complications & ~-7.3 (-22.2, 7.5) & .33 & ~0.06 (-0.1, 0.2) & .45\\
        HCC & ~-1.9 (-4.1, 0.3) & .09 & ~0.01 (-0.01, 0.03) & .47 \\
         \bottomrule
    \end{tabular}
\end{table}

\begin{table}[ht]
    \footnotesize
    \centering
    \caption{Estimates for main effect ($S_k = \emptyset$) or interaction term ($S_k \neq \emptyset$) of sending the current patient to the vertical area on the admission decision and vertical area decision for the next patient. We report $P$ values for Wald tests.\label{tab:results_binary}}
    \begin{tabular}{l l l l l}
        \toprule
        & \multicolumn{2}{c}{\textbf{Admission decision (\%) }} & \multicolumn{2}{c}{\textbf{Vertical area decision (\%)}} \\
        \cmidrule(lr){2-3} \cmidrule(lr){4-5} 
         \textbf{Variable $S_k$} & Interaction (95\% CI) & $P$ & Interaction (95\% CI) & $P$ \\
        \midrule
        $\emptyset$ & ~0.9 (-0.8, 2.7) & .29 &  ~2.0 (0.3, 3.6) & .01 \\
        Congestive heart failure & ~1.6 (-5.5, 8.8) & .65 & -0.7 (-5.5, 4.1) & .76 \\
        Hypertension & -3.7 (-7.4, 0.05) & .05 & ~0.3 (-2.6, 3.3) & .81\\
        Obesity & -0.5 (-5.0, 3.9) & .81 & ~0.1 (-4.1, 4.4) & .94\\
        Diabetes with complications & ~1.0 (-5.2, 7.3) & .75 & -0.1 (-5.2, 4.8) & .95\\
        Diabetes without complications & -4.0 (-11.0, 2.9) & .25 & ~0.6 (-5.2, 6.5) & .83\\
        HCC & -0.1 (-1.1, 0.7) &  .74 &  -0.3 (-1.0, 0.4) & .46 \\
         \bottomrule
    \end{tabular}
\end{table}

Our results suggest sequential bias in the decision to send a patient to a vertical area over a traditional bed. In Table~\ref{tab:results_operational}, we find that sending the current patient to vertical area leads to a significant increase of $0.04$ (95\% CI: $[0.01,0.09]$) tests on average for the next patient. Regarding time to disposition, our findings indicate a marginally significant interaction between the next patient's HCC score and the decision to send the current patient to the vertical area. Although, in general, sending the current patient to vertical area leads to an estimated non-significant increase in average time to disposition of 3.6 minutes for the subsequent patient, this increase is 1.9 (95\% CI: $[-0.3,4.1]$) minutes shorter when the subsequent patient has an HCC score of 1 compared to an HCC score of 0. In Table~\ref{tab:results_binary}, we find that sending the current patient to vertical area leads to a significant increase of 2\% (95\% CI: $[0.3\%,3.6\%]$) in the next patient's probability of being assigned to vertical area. Further, we observe a marginally significant interaction between the next patient's hypertension status and the decision to send the current patient to the vertical area. Although, in general, sending the current patient to vertical area lead to an estimated non-significant increase of 0.9\% in the next patient's admission rate, this rate increase is estimated to be -3.7\% (95\% CI: $[-7.4\%,0.05\%]$) lower when the patient has hypertension compared to when they do not have hypertension. 

\section{Conclusion}\label{sec:bias_discussion}

We introduced a new causal inference framework to examine sequential bias in stochastic service systems from observational data. Sequential bias has been widely observed across diverse domains, including sporting competitions, stock markets, and judging decisions. If we can accurately measure this bias, we can pave the way for enhancing decision-making in stochastic service systems through stochastic modeling and optimization. With this goal in mind, we had to address two obstacles to performing causal inference: interference and a random number of jobs. To handle interference, we drew upon insights from DTRs, building specifically upon the work in \citet{robins_causal_1997,boruvka_assessing_2018,qian_estimating_2021}. Handling the random number of jobs introduced challenges across modeling, identification, and estimation. These challenges include building a causal model that permits a random number of variables, identifying the resulting causal effects, and deriving estimation properties such as double robustness and an efficient score. Our innovative solutions to these challenges represent the core contributions of our work, pushing the boundaries of causal inference forward.

To tackle the issue of modeling a random number of jobs, we conceptualized the decision-making process as a MPP and integrated it within the well-established causal inference framework established by \cite{richardson2013single} and \cite{malinsky2019potential}. By employing the MPP, we were able to effectively account for the number of jobs. However, this required working with an infinite number of random variables, posing certain technical difficulties. Ensuring the integrability of our proposed estimating equations and justifying the exchange between differentiation and an infinite sum in our proof of semiparametric efficiency were among the difficulties we encountered and successfully navigated. These successes not only help us quantify sequential bias, but also offer a way to effectively handle a random number of decisions in a DTR.

Based on the MPP, we introduced two causal effects as measures of sequential bias. These effects differ from their counterpart in the DTR literature \citep{boruvka_assessing_2018, qian_estimating_2021} for their inclusion of variables collected after the intervention in the conditioning set. This inclusion is unavoidable because, in the presence of a random number of jobs, measuring the causal impact that intervening on a current job has on a future job requires that the future job exists. In other words, we always condition on a specific post-intervention variable, namely the binary indicator of whether the future job exists. Moreover, the inclusion of post-intervention variables brings an additional advantage of modeling the future job's outcome using its own characteristics, rather than the characteristics of the current or past job. This approach holds promise for achieving more efficient estimators. Indeed, we derived an efficient score for those specific lag effects that conditioned on \emph{every} post-intervention variable that is a parent of the future job's outcome. Without conditioning on post-intervention variables, our proof would only work for effects on the outcome of the currently intervened job, instead of a future job. This consideration may explain why \citet{qian_estimating_2021}, which did not condition on post-intervention variables, provided an efficient score only in the special case of their causal effect when the outcome was the most immediate.

Nevertheless, it is crucial to acknowledge a caveat when conditioning on post-intervention variables: the interpretation of the lag effect becomes more nuanced. To address the confusion surrounding conditioning on post-intervention variables, \citet{pearl2015conditioning} extensively discusses this topic in a dedicated paper. He highlights the importance of considering whether to condition on the observed value of post-intervention variables or the hypothetical value that these variables would achieve under the contrasting scenarios. These considerations lead to different research questions. In our study, the question was clear: to identify the lag effect, we needed to condition on the hypothetical value rather than the observed value. Consequently, the resulting lag effect is interpreted as an average contrast in a future job's outcome between intervening and not intervening on a current job among those events whose post-treatment variables would achieve a certain value regardless of the intervention. For instance, we may compare the future job's outcome when intervening or not intervening on a current job, focusing on events where the future job exists in either case.

Applying our framework to specific settings comes with several limitations to consider. One such limitation is that, like many other causal inference approaches, we rely on assumptions that cannot be directly verified from data. These assumptions involve considering multiple hypothetical scenarios, of which only one is actually realized. Therefore, the validity of our approach in a given application depends on how well the application can be accurately represented using our proposed causal model. A specific issue to be mindful of, because it could compromise the model's validity, is an unobserved common cause of the future job's outcome and the current job's outcome. Thus, careful consideration of such confounding factors is necessary to ensure the validity of our approach.

Another limitation is the unknown assignment mechanism for the intervention, which needs to be estimated from the available data. Since our estimator is doubly-robust, this implies that if both the probability model of assignment and the outcome model are incorrect, there is a risk of biased estimates. A related issue is the occurrence of intervention assignments with extremely low probabilities, which we then divide by during estimation. The accuracy of these probabilities can impact on estimation. Furthermore, such probabilities suggest that certain jobs are very unlikely to experience one of the intervention conditions, which violates our assumption of positivity. One proposed solution is to exclude cases with extremely low probabilities \citep{lee2011weight}. Alternatively, overlap weights can be used to smoothly reduce the influence of observations with extreme probabilities \citep{li2019addressing}.  Encouragingly, our investigation of double robustness demonstrates that by adding a term in the numerator of our weights related to the probability of assignment, our estimator effectively produces a weighted causal effect similar to one that used overlap weights.

A final limitation is the assumption of non-interference among panels, i.e. that the outcomes within a panel are solely determined by the decisions in that panel, independent of decisions made in other panels. However, this assumption may not hold if the decision-maker for one panel of data were to coordinate their decisions with the decision-maker on another panel. For example, physicians in the ED can potentially impact each other's decisions, making it inappropriate to consider patients treated by different doctors during the same time period as arising from distinct panels.

To conclude our investigation, we estimated lag effects for a specific decision-making scenario in the ED, offering a concrete illustration of how our proposed causal inference framework can be applied. We provide step-by-step guidance on how to implement our method using commonly available statistical software. The estimation of standard errors may be the most challenging aspect, and we include code snippets in the Appendix to facilitate this process. Our analysis reveals a noteworthy finding regarding the influence of a current patient's routing decision on the next patient. Specifically, our data analysis demonstrates that routing the current patient to a vertical area significantly increases both the number of tests performed and the probability of the next patient being routed to the vertical area. These findings provide compelling support for the existence of sequential bias in a setting of paramount importance and profound consequences.

\bibliography{references}
\bibliographystyle{agsm}

\appendix
\input{appendix}

\end{document}

%% file: appendix.tex
\newpage
\setcounter{equation}{0}

{\LARGE \textbf{Appendix}\par}

\section{Regularity Conditions}\label{app:regularity}

To construct the distribution for our MPP, we make use of the Ionescu Tulcea theorem. Letting $\oplus_{k \in \N} {\cal E}_k$ be the product $\sigma$-algebra in $\prod_{k \in \N} \Omega_k$ given a sequence  $(\Omega_k,{\cal E}_k)_{k \in \N}$ of measurable spaces, the statement of the theorem is the following:
\begin{appthm}[Ionescu Tulcea] \label{thm:ionescu_tulcea}
For a sequence of measurable spaces $(\Omega_k,{\cal E}_k)_{k \in \N}$, a probability measure $\PR_0$ on $\Omega_1$, and Markov kernels $\PR_k$ from $\prod_{i=1}^k \Omega_{k}$ to $\Omega_{k+1}$ for $k \in \N$, there exists random variables $Z_k$ taking values in $\Omega_k$ with unique probability distribution $\PR$ on $(\prod_{k \in \N} \Omega_k, \otimes_{k \in \N} {\cal E}_k)$ satisfying the following equation for all $k \in \N$ and $B_i \in {\cal E}_i$:
$$\PR( \{ Z_i \in B_i \}_{i=1}^k ) = \int_{B_1} \PR_0(d \omega_1 ) \int_{B_2} \PR_1(d \omega_2 | \omega_1) \ldots \int_{B_k} \PR_k( d \omega_k | \omega_1,\ldots,\omega_{k-1}).$$
\end{appthm}

In order for the random variables $(T_k,X_k,A_k,Y_k)_{k \in \N}$ to form an MPP, additional conditions on the Markov kernels must be satisfied. These conditions ensure that the times are strictly increasing when they are finite, and increasing otherwise, with the marks taking on the irrelevant mark when the corresponding time is infinite. Specifically, the following assumptions are made:

\begin{appassm}{(Marked point process conditions)}\label{ass:mpp_conditions} The Markov kernels from Assumption~\mref{ass:kernels} satisfy:
    \begin{align*}     \PR^{(T_{k})}\left((t_{k-1}, \infty] \,\big|\, t_{1:k-1},h_{k-1}\right) &= 1 &&\text{if }\,\, t_{k-1} < \infty \\
\PR^{(T_{k})}\left(\{\infty\} \,\big|\, t_{1:k-1},h_{k-1}\right) &= 1 &&\text{if }\,\, t_{k-1} = \infty \\
\PR^{(X_{k})} \left(\mathbb{R}^l \,\big|\, t_{1:k},h_{k-1}\right) &= 1 &&\text{if }\,\, t_{k} < \infty \\\PR^{(X_{k})} \left(\{\Delta\} \,\big|\, t_{1:k},h_{k-1}\right) &= 1 &&\text{if }\,\, t_{k} = \infty.
\end{align*}
and
\begin{align*}\PR^{(A_{k})} \left(\{0,1\} \,\big|\, x_{k},h_{k-1}\right) &= \PR^{(Y_{k})} \left(\mathbb{R} \,\big|\, x_{k},a_{k},h_{k-1}\right) = 1 &&\text{if }\,\, x_{k} \neq \Delta \\\PR^{(A_{k})} \left(\{\Delta\} \,\big|\, x_{k},h_{k-1}\right) &=  \PR^{(Y_{k})} \left(\{\Delta\} \,\big|\, x_{k},a_{k},h_{k-1}\right) = 1 &&\text{if }\,\, x_{k} = \Delta.
\end{align*}
\end{appassm}

For estimation, we need several assumptions for the proposed estimator to be integrable (Lemma~\mref{lem:integrability}), consistent (Theorem~\mref{thm:asymp_consistency}), and asymptotically normal (Theorem~\mref{thm:asym_normality}). The next assumption ensures that $U_{\rm{stacked}}$ is integrable. 

\begin{appassm}{(Integrability  conditions)}\label{ass:int_regularity}
The random number of jobs has finite expectation, viz, $\E[K] < \infty$. We also have the following bound:
$$\E\left[  \begin{Vmatrix} \begin{bmatrix} U_k \\ L_k \\ M_k  \end{bmatrix} \end{Vmatrix} \mathbbm{1}_{\{k \leq K\}} \right] \leq C \PR(k \leq K)$$ 
for some constant $C$.
\end{appassm}

In addition to assumptions above, we also invoke the assumptions below for consistency and asymptotic normality. The following assumptions for consistency incorporate the assumptions in Lemma 1 for \citet{tauchen1985diagnostic}, with the exception that the separability condition is replaced with a more direct condition on measurability of a function involving a supremum: 

\begin{appassm}{(Consistency conditions)} \label{ass:cons_regularity}
In addition to the conditions needed for integrability (Assumption~\ref{ass:int_regularity}), the set $\Theta$ is compact, $\E[U_{\rm{stacked}}]$ has a unique root $\theta_{\infty}$ in $\Theta$, and almost surely $\E_n[U_{\rm{stacked}}]$ has a root $\theta_n$  in $\Theta$. Take $Z:=(X_k,A_k,Y_k)_{k \in \{1,\ldots,K\}}$. Let $\psi$ be the mapping from $\theta$ and $Z$ to $\psi(Z,\theta)=U_{\rm{stacked}}$ and assume $\psi$ is
\begin{itemize}
    \item  almost surely continuous for each $\theta \in \Theta$
    \item almost surely bounded above by a measurable function $d(z)$ that is invariant to $\theta$ and has finite mean
    \item measurable for each $\theta \in \Theta$
    \item is defined such that 
    $$ \sup_{ \theta \in \Theta: \begin{Vmatrix} \theta - \theta^* \end{Vmatrix} < \delta} \begin{Vmatrix} U_{\rm{stacked}} -U_{\rm{stacked}} \big|_{\theta = \theta^*} \end{Vmatrix}$$
    is measurable for all $\delta > 0$ and $\theta^* \in \Theta$.
    \end{itemize}
\end{appassm}
The hardest step in the proof of consistency (Theorem~\ref{thm:asymp_consistency}) is showing a uniform law of large numbers from these conditions, for which we refer the reader to the proof of Lemma 1 in \citet{tauchen1985diagnostic}. Asymptotic normality (Theorem~\mref{thm:asym_normality}) requires yet stronger conditions, which are given as followed:

\begin{appassm}{(Asymptotic normality conditions)} \label{ass:amvn_regularity}
In addition to the conditions needed for integrability (Assumption~\ref{ass:int_regularity}) and consistency (Assumption~\ref{ass:cons_regularity}), we have
\begin{itemize}   
 \item Measurable $\dot{\psi}(z)$ exists with $\E[\dot{\psi}(Z)^2] < \infty$ such that for $\theta_1,\theta_2$ in $\Theta$ and in a neighborhood of $\theta_{\infty}$
 $$\begin{Vmatrix} \psi(z,\theta_1) - \psi(z,\theta_2) \end{Vmatrix} \leq \dot{\psi}(x) \begin{Vmatrix} \theta_1 - \theta_2 \end{Vmatrix}$$
    \item The matrices
\begin{align*}
\mathbb{B} = \E\left[ \nabla_{\theta}U_{\rm{stacked}} \big|_ {\theta = \theta_{\infty}}\right]; \qquad 
\mathbb{C} = \E\left[U_{\rm{stacked}}U_{\rm{stacked}}' \big|_ {\theta = \theta_{\infty}} \right],
\end{align*}
exist and are invertible.
\end{itemize}
\end{appassm}

\section{Proof of Integrability (Lemma~\mref{lem:integrability})} \label{app:integrability}

With the assumptions for integrability stated in the last section, we are ready to provide the proof of integrability (Lemma~\mref{lem:integrability}):

\Pf 
Our proof is for $U$. The same argument holds for $L$ and $M$.
Note that the triangle inequality implies that $\begin{Vmatrix} U \end{Vmatrix} \leq \sum_{k=1}^K \begin{Vmatrix} U_k \end{Vmatrix}$ almost surely.  As a result, we have that
$$\E\left[ \begin{Vmatrix} U \end{Vmatrix} \right] = \E\left[ \begin{Vmatrix}\sum_{k=1}^K U_k \end{Vmatrix} 
\right]\leq \E\left[ \sum_{k=1}^K \begin{Vmatrix} U_k \end{Vmatrix} \right].$$

Next note that we can write $\E\left[ \sum_{k=1}^K \begin{Vmatrix} U_k\end{Vmatrix} \right]$ as $\E\left[ \sum_{k=1}^{\infty} \begin{Vmatrix} U_k \end{Vmatrix} \mathbbm{1}_{\{k \leq K\}} \right]$.  Applying the monotone convergence theorem to this last expression allows us to write 
$\E\left[ \sum_{k=1}^{\infty} \begin{Vmatrix} U_k \end{Vmatrix} \mathbbm{1}_{\{k \leq K\}} \right] = \sum_{k=1}^{\infty} \E\left[  \begin{Vmatrix} U_k \end{Vmatrix} \mathbbm{1}_{\{k \leq K\}} \right]$.
The second part of Assumption~\ref{ass:int_regularity} is that  
\begin{align*}
    \E\left[  \begin{Vmatrix} U_k \end{Vmatrix} \mathbbm{1}_{\{k \leq K\}} \right] \leq C \PR(k \leq K),
\end{align*}
which implies
\begin{align*}
  \sum_{k=1}^{\infty} \E\left[  \begin{Vmatrix} U_k \end{Vmatrix} \mathbbm{1}_{\{k \leq K\}} \right] \leq  C \sum_{k=1}^{\infty}  \PR(k \leq K) = C \E[ K].
\end{align*}
The first part of Assumption~\ref{ass:int_regularity} 
 gives the finiteness of $\E[K]$, which completes the proof. 
\EndPf

\section{Proof of Double Robustness (Theorem~\mref{thm:double_robust})} \label{app:double_robust}

Here, we provide the proof for Theorem~\mref{thm:double_robust}, which relates our estimator to $\beta_{\infty}$ to our target effect $\zeta_{\rm{marg}}(k,\ell,S_k)$ in situations when some of our models are correct.

\Pf
The proof of this theorem involves understanding what it means for $\theta_{\infty}$ to be a solution for $\E[ U_{\rm{stacked}} ].$ We accomplish this in several steps.

\medskip

\textbf{Step 1: Re-write the estimating equation.} We first note that
$$\E[ U_{\rm{stacked}} ] = \E\left[ \sum_{k=1}^{K-\ell} U_k \right] = \E\left[ \sum_{k=1}^{\infty} U_k \mathbbm{1}_{\{ k \leq K \}} \right] = \sum_{k=1}^{\infty} \E\left[ U_k | T_k < \infty\right] \PR( T_{k+\ell} < \infty).$$
We were able to swap the infinite sum with the expectation by utilizing the conditions for integrability (Assumption~\ref{ass:int_regularity}) and the dominated convergence theorem. In particular, we established the finiteness of the expectation of $\begin{Vmatrix} U_{\rm{stacked}} \end{Vmatrix}$ in Lemma~\mref{lem:integrability}, which dominates $U_{\rm{stacked}}.$ Further, the solution $\theta_{\infty}$ to the estimating equation:
$$\E[ U_{\rm{stacked}} ]=\sum_{k=1}^{\infty} \E\left[  W_k\left(Y_{k+\ell} - g(R_k)'\alpha - A_k f(R_k)'\beta \right) \begin{bmatrix} g(R_k) \\ A_k f(S_k) \end{bmatrix} \bigg| T_{k+\ell} < \infty \right] \PR( T_{k+\ell} < \infty)=0$$
is also a solution to
$$ \sum_{k=1}^{\infty} \E\left[ W_k\left(Y_{k+\ell} - g(R_k)'\alpha - A_k f(S_k)'\beta \right) (A_k - q_k(S_k; \xi))f(S_k)\,\big|\, T_{k+\ell} < \infty \right] \PR( T_{k+\ell} < \infty)=0.$$
This can be shown by subtracting the rows of $U_k$ involving $q_k(S_k; \xi)f(S_k)$ in $g(R_k)$ from the rows involving $A_k f(S_k),$ which is a direct consequence of our requirement that $g(R_k)$ includes $q_k(S_k; \xi)f(S_k).$

\medskip

\textbf{Step 2: Simplify the new estimating equation.} We want to show that
\begin{equation}
    \begin{aligned}
        &\E \left[ W_k \left(Y_{k+\ell} - g(R_k)'\alpha- A_k f(S_k)'\beta\right)(A_k -q_k(S_k; \xi))f(S_k) | R_k, T_{k + \ell} < \infty \right] \\
    &= 
    q_k(S_k;\xi)(1-q_k(S_k;\xi))f(S_k)\left( \zeta(k,\ell,R_k) - f(S_k)' \beta\right) + \epsilon_k 
    \end{aligned}\label{eq:double_step1}
\end{equation}
for some (error) term $\epsilon_k$. To that end, let $Z_k$ be shorthand for $Y_{k+\ell}-g(R_k)'\alpha-A_k f(R_k)'\beta$, and $\rho_k(R_k)$ shorthand for $\E[A_k | R_k, T_{k+\ell} < \infty].$ Notice that
\begin{align*}
    &\E\left[ W_k Z_k (A_k - q_k(S_k; \xi))f(S_k)| R_k, T_{k+\ell} < \infty \right] \\ 
    &\qquad=\E\left[ W_k Z_k (A_k - q_k(S_k; \xi))f(S_k) | R_k, A_k=1,  T_{k+\ell} < \infty \right] \rho_k(R_k) + \\
    &\qquad\quad\,\E\left[ W_k Z_k (A_k - q_k(S_k; \xi))f(S_k) | R_k, A_k=0, T_{k+\ell} < \infty \right] (1-\rho_k(R_k)) \\
    &\qquad= \E\left[ Z_k | R_k, A_k=1, T_{k+\ell} < \infty \right] \frac{q_k(S_k;\xi)}{p_k(R_k;\eta)} (1-q_k(S_k;\xi)) f(S_k) \rho_k(R_k) + \\
    &\quad\qquad\,\E\left[ Z_k | R_k, A_k=0, T_{k+\ell} < \infty \right] \frac{1-q_k(S_k;\xi)}{1-p_k(R_k;\eta)}(0-q_k(S_k;\xi))f(S_k) (1-\rho_k(R_k)), 
\end{align*}
which we re-arrange to:
\begin{align*}
 &q_k(S_k;\xi)(1-q_k(S_k;\xi))f(S_k)\left( \E\left[ Z_k | R_k, A_k=1, T_{k+\ell} < \infty \right] - \E\left[ Z_k | R_k, A_k=0, T_{k+\ell} < \infty \right] \right) + \epsilon_k
\end{align*}
where $\epsilon_k$ is the remaining term:
\begin{align*}
    &q_k(S_k;\xi)(1-q_k(S_k;\xi))f(S_k)(\rho_k(R_k)-p_k(R_k;\eta))\left( \frac{\E\left[ Z_k | R_k, A_k=1, T_{k+\ell} < \infty \right]}{p_k(R_k;\eta)} + \frac{\E\left[ Z_k | R_k, A_k=0, T_{k+\ell} < \infty \right]}{1-p_k(R_k;\eta)} \right).
\end{align*}
Focusing on
$$\E\left[  Z_k \big| R_k, A_k=1, T_{k + \ell} < \infty \right] - \E\left[  Z_k \big| R_k, A_k=0, T_{k + \ell} < \infty \right],$$
we use the definition of $Z_k$ and Lemma~\mref{lem:identification} to simplify this expression to
\begin{align*}
    &\E\left[  Y_{k+\ell}\big| R_k, A_k=1, T_{k + \ell} < \infty \right] - \E\left[  Y_{k+\ell} \big| R_k, A_k=0, T_{k + \ell} < \infty \right] - f(R_k)'\beta \\
    &= \zeta(k,\ell,R_k) - f(R_k)'\beta.
\end{align*}
Upon substituting this expression into the one above, we arrive at Equation~\ref{eq:double_step1}.

\medskip

\textbf{Step 3: Show $\epsilon_k=0$.} We want to show that $\epsilon_k$ is zero under the conditions of the theorem. We consider the two cases. First assume the conditional probability model in the denominator of weights (Equation~\mref{eq:denominator_model}) is correct for $\theta_{\infty} \in \Theta$ so that $\rho_k(R_k) = p_k(R_k;\eta_{\infty}).$ Then, it is clear that $\epsilon_k$ evaluated at $\theta_{\infty}$ is zero. Alternatively, assume the outcome model is correct (Equation~\mref{eq:outcome_model}) for $\theta_{\infty} \in \Theta$ so that
\begin{align*}
    \E[ Y_{k+\ell} | R_k, A_k, T_{k+\ell} < \infty ] = g(R_k)'\alpha_{\infty} + A_k f(S_k)'\beta_{\infty}.
\end{align*}
Then,
\begin{align*}
\E[ Y_{k+\ell} - g(R_k)'\alpha_{\infty} - A_k f(S_k)'\beta_{\infty} | R_k, A_k, T_{k+\ell} < \infty ] = 0
\end{align*}
and again, it is clear that $\epsilon_k$ evaluated at $\theta_{\infty}$ is zero.

\medskip

\textbf{Step 4: Show $\mathbb{G} \beta_{\infty} = g$.}  From above, we know the conditions of the theorem imply that, if we were to evaluate
$$ \sum_{k=1}^{\infty} \E\left[ W_k\left(Y_{k+\ell} - g(R_k)'\alpha - A_k f(R_k)'\beta \right) (A_k - q_k(S_k; \xi))f(S_k)\,\big|\, T_{k+\ell} < \infty \right] \PR( T_{k+\ell} < \infty)=0$$
at $\theta_{\infty}$ and apply the law of iterated expectation conditioning on $R_t$, we would get
$$ \sum_{k=1}^{\infty} \E\left[q_k(S_k;\xi_{\infty})(1-q_k(S_k;\xi_{\infty}))\left( \zeta(k,\ell,R_k) - f(S_k)' \beta_{\infty}\right)f(S_k) \big|\, T_{k+\ell} < \infty \right]\PR( T_{k+\ell} < \infty)=0.$$
Hence,
\begin{align*}
&\sum_{k=1}^{\infty} \E\left[ q_k(S_k;\xi_{\infty})(1-q_k(S_k;\xi_{\infty}))f(S_k)\zeta(k,\ell,R_k)\big|\, T_{k+\ell} < \infty \right]\PR( T_{k+\ell} < \infty)\\
&= \left( \sum_{k=1}^{\infty} \E\left[ q_k(S_k;\xi_{\infty})(1-q_k(S_k;\xi_{\infty}))f(S_k) f(S_k)' \big|\, T_{k+\ell} < \infty \right]\PR( T_{k+\ell} < \infty) \right) \beta_{\infty}.
\end{align*}
This last term is $\mathbb{G} \beta_{\infty}$ where $\mathbb{G}$ is given in the statement of the theorem. One more application of law of iterated expectation, this time conditioning on $S_t$, and the definition of $\zeta_{\rm{marg}}(k,\ell,S_k)$ means
\begin{align*}
&\E\left[ q_k(S_k;\xi_{\infty})(1-q_k(S_k;\xi_{\infty}))f(S_k)\zeta(k,\ell,R_k)\Big|\, T_{k+\ell} < \infty \right] \\
&= \E\left[ q_k(S_k;\xi_{\infty})(1-q_k(S_k;\xi_{\infty}))f(S_k)\E\left[\zeta(k,\ell,R_k) \big|\, S_k, T_{k+\ell} \right]\Big|\, T_{k+\ell} < \infty \right] \\
&= \E\left[ q_k(S_k;\xi_{\infty})(1-q_k(S_k;\xi_{\infty}))f(S_k) \zeta_{\rm{marg}}(k,\ell,S_k)\Big|\, T_{k+\ell} < \infty \right],
\end{align*}
which is just the vector $g$ in the statement of the theorem. Thus, we have $\mathbb{G} \beta_{\infty} = g.$

\medskip

\textbf{Step 5: Show $\beta_{\infty} = \beta_*$.} Our last step is to assume that, in addition, $\mathbb{G}$ is invertible and the model of the lag effect is correct (Equation~\mref{eq:lag_model}) for some $\theta_* = (\xi_*,\eta_*,\alpha_*,\beta_*) \in \Theta$. In this case, 
$f(S_k)'\beta_* = \zeta_{\rm{marg}}(k,\ell,s)$ so that $\mathbb{G} \beta_{\infty} = g$ becomes
$$\mathbb{G} \beta_{\infty} = \mathbb{G} \beta_*.$$
Invertibility of $\mathbb{G}$ ensures $\beta_{\infty} = \beta_*$, thus completing the proof.
\EndPf

\section{Proof of Efficient Score (Theorem~\mref{thm:opt_efficiency})} \label{app:efficiency}

In this section, we prove Theorem~\mref{thm:opt_efficiency} which provides the efficient score for $\beta$. This theorem illuminates the motivation behind our final estimator. Before we get started, we want to mention that our approach is not self-contained, relying heavily on concepts such as Hilbert spaces, tangent spaces, and parametric submodels. It is influenced by concepts presented in the textbook by \citet{tsiatis_semiparametric_2006}, as well as a recent paper by \citet{qian_estimating_2021} and its  antecedent by \citet{robins_correcting_1994}.  We have adapted the methods from these sources to fit our specific research question, which involves dealing with a random number of jobs, having certain assumptions on parents, and conditioning on post-intervention variables.  

Efficiency is expressed relative to a Hilbert space ${\cal H}$ and a semi-parametric model ${\cal M}$. If $\PR_0$ denotes the correct distribution and $\E_0$ expectation with respect to $\PR_0$, then the Hilbert space ${\cal H}$ consists of $m$-dimensional measurable functions of $(T_i,X_i,A_i,Y_i)_{i=1}^{\infty}$ with mean zero and finite variance with respect to $\PR_0$. Functions in this space are uniquely identified up to sets of probability ($\PR_0$) zero. The inner product of two functions in ${\cal H}$ is the expectation of their dot product under $\E_0$.

The semi-parametric model ${\cal M}$ is based on our original causal model. We first change variables from $Y_{k+\ell}$ to 
$${{Q}}_{k+\ell} = Y_{k + \ell} - \mathbbm{1}_{\{T_{k+\ell} < \infty\}} \left( \E_0\left[ Y_{k+\ell} | A_k=0, R_k, T_{k+\ell} < \infty \right] + A_k f(R_k)'\beta \right)$$
for $k \in N$ and from $Y_k$ to ${{Q}}_k = Y_k$ for $k \leq \ell$.
The relation
\begin{align*}
    f(R_k)'\beta = \E[Y_{k+\ell} | A_k=1, R_k, T_{k+\ell}<\infty] -\E[Y_{k+\ell} | A_k=0, R_k, T_{k+\ell} < \infty]
\end{align*}
then implies
\begin{align*}
0 = \E[{{Q}}_{k+\ell} | A_k=1, R_k, T_{k+\ell}<\infty] -\E[{{Q}}_{k+\ell} | A_k=0, R_k, T_{k+\ell} < \infty],
\end{align*}
when $\E$ denotes expectation with respect to an arbitrary distribution $\PR$. Additionally, we note
\begin{equation} \label{eq:kappa_centered}
\begin{aligned}
   &\E_0[ {{Q}}_{k+\ell} | \Pa({{Q}}_{k+l}), T_{k+\ell} < \infty] \\
   &=\quad\E_0[ Y_{k+\ell} | A_k, R_k, T_{k+\ell} < \infty] -  \E_0\left[ Y_{k+\ell} | A_k=0, R_k, T_{k+\ell} < \infty \right] + A_k f(R_k)'\beta =0. 
\end{aligned}
\end{equation}
Further, the parents of ${{Q}}_{k+\ell}$ are the parents of $Y_{k+\ell}$ except with $Y_{i + \ell}$ replaced by ${{Q}}_{i+\ell}$. 


\begin{appdefn}{(Semi-parametric model)} \label{def:modeling_space}
Fix $\ell \in \mathbb{R}$, a deterministic function $f$, and $R_k=\Pa(Y_{k+\ell})\setminus A_k.$ Consider a collection ${\cal M}$ of probability distributions for the random variables $(T_i,X_i,A_i,Y_i)_{i=1}^{\infty}.$ Each $\PR \in  {\cal M}$ is constructed in two steps. We first construct $\PR$ from Markov kernels as in Assumption~\mref{ass:probability_model} except that we replace the $Y_i$ by the $Q_i$. We then construct the $Y_i$ using the transformation above for some $\beta$ in an open set $B$ of $\mathbb{R}^m.$
Further, assume
\begin{itemize}

    \item For all $k \in \N$ and $\PR \in {\cal M}$,    \begin{align}\label{eq:restricted_moment2}
    0 = \E[{{Q}}_{k+\ell} | A_k=1, R_k, T_{k+\ell}<\infty] -\E[{{Q}}_{k+\ell} | A_k=0, R_k, T_{k+\ell} < \infty],
    \end{align}
    where expectation is taken with respect to $\PR$. 
    \item Each $\PR \in {\cal M}$ is absolutely continuous with respect to the same measure $\mu$. We denote the associated density (Radon-Nikodym derivative) as $p$. 
    \item The correct model $\PR_0$ is in  ${\cal M}.$ We denote the associated density by $p_0$, Markov kernels by $\PR^{(Z)}_0$, and $m$-dimensional parameter in Equation~\eqref{eq:restricted_moment2} by $\beta_0$.
    \item For distributions $ \PR_{\beta} \in {\cal M}$ with  associated density $p_{\beta},$ constructed from $\beta$ and Markov kernels $\PR_{0}^{(Z)} \in {\cal M},$ the function $\log p_{\beta}$ is continuously differentiable with respect to $\beta$ when evaluated at $\beta=\beta_0$. The ensuing score function $S_{\beta} =\nabla_{\beta} \log p_{\beta} |_{\beta=\beta_0}$ exists in ${\cal H}.$
    \end{itemize}
\end{appdefn}

\smallskip

With these preparations in place, the general flow of the proof can be outlined as follows. Initially, we define the nuisance tangent space $\Gamma$ associated with ${\cal M}$. Next, our focus shifts towards determining the score function $S_{\beta}$ pertaining to our target of interest, namely $\beta$. Subsequently, we decompose $S_{\beta}$ into two components: $S_{\beta}-S_{\beta}^{\perp}$, which resides within $\Gamma$, and $S_{\beta}^{\perp}$, which exists in the orthogonal complement of $\Gamma$, denoted by $\Gamma^{\perp}$. Utilizing the uniqueness of this decomposition, we establish that $S_{\beta}^{\perp}$ represents the projection of $S_{\beta}$ onto the orthogonal complement of $\Gamma$. As the efficient score function $S_{\rm{eff}}$ is obtained by projecting $S_{\beta}$ onto the orthogonal complement of $\Gamma$, we deduce that $S_{\rm{eff}}$ is equivalent to $S_{\beta}^{\perp}$. We will now delve into each of these steps individually.

\textbf{Step 1. Characterizing the  nuisance tangent space $\Gamma$.} 

For the next lemma, it is easier to characterize $\Gamma$ if we relabel each variable as $(Z_1,Z_2,Z_3,\ldots)$. The relabelling does not matter as long as no variable precedes its parents in the list. We introduce the set $\mathbb{K} \subset \N$ to capture the indices of the $Z_i$ that correspond to ${{Q}}_{k+\ell}$ for some $k \in N$. For instance, if $Z_4={{Q}}_{1+\ell}$, then $4 \in K$. We use $\Pa(Z_i)$ to denote the parents of $Z_i$. For example, if $Z_1=T_1$ and $Z_2=X_1$, then $\Pa(Z_2)=\Pa(X_1)=T_1=Z_1.$ Last, for each $k \in \mathbb{K},$ we introduce $\sigma$-algebras $J_{k,1}$ and $J_{k,0}$ to capture the conditioning events in the updated moment restriction (Equation~\ref{eq:restricted_moment2}). If $Z_4 = {{Q}}_{1+\ell},$ then
$J_{4,1}$ is the $\sigma$-algebra induced from $A_1=1$, $R_1$, and $T_{1+\ell}<\infty$, and $J_{4,0}$ is the $\sigma$-algebra induced from $A_1=0$, $R_1$, and $T_{1+\ell}<\infty.$ In other words,  $J_{k,1}$ and $J_{k,0}$ are defined such that for all $k \in \mathbb{K}$
\begin{align*}
    \E_0\left[ Z_k | J_{k,1} \right] -\E_0 \left[ Z_k | J_{k,0} \right] = 0.
\end{align*}
Although not explicitly stated in the following derivations, we heavily rely on the assumption that $J_{k,1}$ and $J_{k,0}$ are subsets of the sigma-algebra induced by $\Pa(Z_k)$. To understand and verify this assumption, consider that $\Pa({{Q}}_{k+\ell})=(A_k, R_k)$. While $T_{k+\ell}$ is not included in $\Pa({{Q}}_{k+\ell})$, the event $\{T_{k+\ell} < \infty\}$ is equivalent to $\{X_{k+\ell} \neq \Delta\}$, where $X_{k+\ell}$ is included in $\Pa({{Q}}_{k+\ell})$. This equivalence is based on the construction of MPP.

\begin{applem}{(Nuisance tangent space)} \label{lem:gamma}
    The nuisance tangent space $\Gamma$ with respect to ${\cal M}$ and the inferential target $\beta_0$ is 
    \begin{align*}
\oplus_{k=1}^{\infty} {\cal H}_k = \left\{ \sum_{k=1}^{\infty} h_k : h_k \in {\cal H}_k, \sum_{k=1}^{\infty} \E_0\left[ h_k \cdot h_k \right] < \infty\right\},
\end{align*}
    where, if $k \in \mathbb{K}$, ${\cal H}_k$ is
    $$\left\{ h_k(Z_k,\Pa(Z_k)) \in {\cal H} : \E_0[h_k \, | \, \Pa(Z_k)] = \E_0[h_k Z_k \, | \, J_{k,1}] - \E_0[h_k Z_k \, | \, J_{k,0}] = 0 \right\}$$
 and, otherwise, is
$$\big\{ h_k(Z_k,\Pa(Z_k)) \in {\cal H} : \E_0[h_k \, | \, \Pa(Z_k)] = 0 \big\}.$$
\end{applem}

\Pf 
To prove the lemma, first note that ${\cal H}_k$ are closed, orthogonal linear subspaces of ${\cal H}$. To see that they are orthogonal, take any $h_i(Z_i,\Pa(Z_i)) \in {\cal H}_i$ and $h_j(Z_j,\Pa(Z_j))\in {\cal H}_j$ with $i < j$. Hereafter, we suppress the dependence of $h_i,h_j$ on its arguments when the arguments can be inferred from context. Then,
\begin{align*}
    \E_0[ h_i \cdot h_j ] &= \E_0\left[ \E_0[ h_i \cdot h_j | Z_{1:j-1}] \right] && \text{\it Law of iterated expectations}\\
    &=\E_0\left[ h_i \cdot \E_0[ h_j | Z_{1:j-1}] \right] && \text{\it  $h_i$ is a function of $Z_{1:j-1}$} \\
    &=\E_0\left[ h_i \cdot \E_0[ h_j | \Pa(Z_j)] \right] && \text{\it  $h_j \ind Z_{1:j-1} \setminus \Pa(Z_j) | \Pa(Z_j)$}  \\
    &=\E_0\left[ h_i \cdot 0 \right] && \text{\it  Definition of ${\cal H}_j$} \\
    &=0.
\end{align*}


We first show the inclusion $\oplus_{k=1}^{\infty} {\cal H}_k \subseteq \Gamma.$ Consider any $h \in \oplus_{k=1}^{\infty} {\cal H}_k$ so that $$h = \sum_{k=1}^{\infty} h_k$$ where $h_k \in {\cal H}_k.$ We want $h \in \Gamma.$ It suffices to consider almost surely bounded $h$, since arbitrary $h \in \oplus_{k=1}^{\infty} {\cal H}_k$ can be expressed as a limit in ${\cal H}$ of almost surely bounded functions in $\oplus_{k=1}^{\infty} {\cal H}_k$, and $\Gamma$ contains by definition its limit points.

Fix $j \in N$ and consider the partial sum $\sum_{k=1}^j h_k \in {\oplus_{k=1}^{\infty} {\cal H}_k}.$ We construct distributions $\PR$ from $\beta=\beta_0$ and from Markov kernels given by
\begin{align*}
    \PR^{(Z_k)}(D |\Pa(Z_k)) = \begin{cases}  \PR_0^{(Z_k)}(D |\Pa(Z_k)) + \E_0[ \mathbbm{1}_{D} \gamma_k'h_k | \Pa(Z_k) ]& k=1,\ldots,j \\ \PR_0^{(Z_k)}(D |\Pa(Z_k)) & k > j
    \end{cases}
\end{align*}
for measurable sets $D$ and for $\gamma_k \in \mathbb{R}^m$ restricted to a sufficiently small set so that $1+\gamma_k' h_k > 0$ for $k=1,\ldots,j$. That we can find such small $\gamma_k$ follows from the boundedness of $h$. We argue that distributions $\PR$ so constructed form a parametric sub-model. To check this, we must show they live in ${\cal M}$ and contain $\PR_0$. The latter is clear, by taking $\gamma_k=0.$ The former is clear for $k > j$, because the Markov kernels $\PR^{(Z_k)}$ are simply $\PR_0^{(Z_k)}$ and because $\beta=\beta_0$, and follows for $k \leq j$ by noting two things. The first is that we are indeed working with well-defined Markov kernels, since $\PR^{(Z_k)}(D |\Pa(Z_k)) \in [0,1]$ for measurable sets $D$ and
\begin{align*}
\PR^{(Z_k)}({\cal X}^{(Z_k)} |\Pa(Z_k)) &= 
\PR_0^{(Z_k)}({\cal X}^{(Z_k)} |\Pa(Z_k)) + \E_{0}\left[ \mathbbm{1}_{{\cal X}^{(Z_k)}}\gamma_k'h_k | \Pa(Z_k) \right] \\
&=  1 + \gamma_k'\E_{0}\left[ h_k | \Pa(Z_k) \right] \\
&=1.
\end{align*}
The second is that for $k \in \mathbb{K} \cap \{1,\ldots, j\},$
\begin{align*}
&\E[ Z_k | J_{k,1} ] - \E[ Z_k | J_{k,0} ] \\ 
&\quad = \E_0[ Z_k | J_{k,1} ] - \E_0[ Z_k | J_{k,0} ] + \gamma_k'\left( \E_0[ h_k Z_k | J_{k,1} ] - \E_0[ h_k Z_k | J_{k,0} ]  \right) && \text{\it Definition of the distributions }\\
&\quad= 0 + \gamma_k'\left( \E_0[ h_k Z_k | J_{k,1} ] - \E_0[ h_k Z_k | J_{k,0} ]  \right) && \text{\it Definition of $\PR_0$ and Equation~\ref{eq:restricted_moment2} }\\
&\quad = 0.  && \text{\it Definition of ${\cal H}_k$ and $h_k \in {\cal H}_k$ }
\end{align*}
We conclude that we indeed have a parametric submodel of ${\cal M}.$

For this particular submodel, the score function with respect to the nuisance parameters $\gamma_k$ is  $\begin{bmatrix} h_1' & \ldots & h_j' \end{bmatrix}.'$
If we take $p_{\gamma_1,\ldots,\gamma_j}$ to be the density of the distribution in our parametric submodel with nuisance parameters $\gamma_1,\ldots,\gamma_j$, then the form of the score function follows from the observation
\begin{align*}
\nabla_{\gamma_k} \log p_{\gamma_1,\ldots,\gamma_j}(Z_{1:\infty}) \Big|_{\gamma_k = 0} = h_k.
\end{align*}
Any product of a matrix $B$ (with $m$ rows) and this score function are, by definition, in the nuisance tangent space $\Gamma.$ In particular, if we take $B=\begin{bmatrix} \mathbb{I} & \ldots & \mathbb{I} \end{bmatrix}$ with $\mathbb{I}$ the $m$-by-$m$ identity matrix, then we see that the partial sum $\sum_{k=1}^j h_k$ lives in $\Gamma.$ Further, any limit points in the Hilbert space ${\cal H}$ of scores functions of parametric submodels of ${\cal M}$ are also, by definition, in $\Gamma.$ Therefore, $\lim_j \sum_{k=1}^j h_k = \sum_{k=1}^{\infty} h_k = h$ is also in $\Gamma.$ This shows every bounded function $h \in \Gamma$ and hence even arbitrary $h \in \oplus_{k=1}^{\infty} {\cal H}_k$ lives in $\Gamma$, upon which we conclude $\oplus_{k=1}^{\infty} {\cal H}_k \subseteq \Gamma.$

\bigskip

It remains to show the reverse inclusion: $\Gamma \subseteq \oplus_{k=1}^{\infty} {\cal H}_k.$ It will suffice to show that $B h \in \oplus_{k=1}^{\infty} {\cal H}_k$ for any matrix $B$ (with $m$ rows) and a nuisance score function $h$ of a parametric submodel of ${\cal M}$. This is sufficient, since arbitrary $g \in \Gamma$ can be expressed as the limit in ${\cal H}$ of such products, and $\oplus_{k=1}^{\infty} {\cal H}_k$ is closed and linear, and hence contains any of such limits. We consider parametric submodels of ${\cal M}$ that satisfy the following:
\begin{itemize}
\item Each distribution $\PR_{\gamma,\beta}$ in the submodel is constructed from $\beta$ and Markov kernels $\PR_{\gamma}^{(Z_k)}$ that depend on a finite-dimensional real-valued vector $\gamma.$ 
\item $\PR_{\gamma_0,\beta_0} = \PR_0$
for some $\gamma_0.$ 
\item $\log p_{\gamma,\beta_0}$ is continuously differentiable with respect to $\gamma$, where $p_{\gamma,\beta}$ is the density associated with $\PR_{\gamma,\beta}$. 
\item $B h \in {\cal H}$ for real-valued matrix $B$, where $h=\nabla_{\gamma} \log p_{\gamma,\beta_0} |_{\gamma=\gamma_0}$ is the nuisance score function. 
\end{itemize}

Our first challenge is to argue that 
$$h = \nabla_{\gamma} \log p_{\gamma,\beta_0} |_{\gamma=\gamma_0} = \sum_{k=1}^{\infty} S_{\gamma,k}, $$
where
\begin{align*}
S_{\gamma,k} = \nabla_{\gamma} \log p_{\gamma,\beta_0}(Z_k | \Pa(Z_k)) \Big|_{\gamma=\gamma_0}
\end{align*}
and $p_{\gamma,\beta}(Z_k | \Pa(Z_k))$ is the relevant conditional density determined from joint density $p_{\gamma,\beta}.$ While expressing $h$ as an infinite sum may appear to pose a technical difficulty, it actually simplifies to a finite sum almost surely. We can see this simplification by examining $p_{\gamma,\beta}(z_{1:\infty})$ for a specific sequence $z_{1:\infty}$ belonging to ${\cal X}^{(Z_{1:\infty})}.$ We can narrow our focus to those $z_{1:\infty}$ that have a special property: there exists an index $n \in N$ such that the elements of $z_{n:\infty}$ take the form $(\infty,\Delta,\Delta,\Delta,\infty,\Delta,\Delta,\Delta,\ldots)$. This selection is appropriate because, for all distributions in ${\cal M}$, the count $K$ of non-infinite times in our MPP is almost surely finite, and once we encounter an infinite time point in the MPP, all subsequent times and marks become, almost surely, $\infty$ and the irrelevant mark $\Delta$. Hence, if a given $z_{1:\infty}$ does not satisfy the aforementioned property, then the densities $p_{\gamma,\beta}(z_{1:\infty})$ and $p_0(z_{1:\infty})$ are both zero for any $\gamma$ and $\beta$. For the same reason above (i.e. once we encounter an infinite time the remaining sequence is determined), it also follows that, for such $z_{1:\infty},$ the conditional density $p_{\gamma,\beta}(z_{n+1:\infty}|z_{1:n})$ is identically equal to $p_0(z_{n+1:\infty}|z_{1:n})$ for any $\gamma$ and $\beta$. Similarly, the conditional density $p_{\gamma,\beta}(z_{k+1}|z_{1:k})$ is identically equal to $p_0(z_{k+1}|z_{1:k})$ for any $k>n$ and any $\gamma$ and $\beta$. This affords us the decomposition:
\begin{align*}
    \nabla_{\gamma} \log p_{\gamma,\beta_0}(z_{1:\infty}) = \nabla_{\gamma} \log p_{\gamma,\beta_0}(z_{1:n}) + \nabla_{\gamma} \log p_{\gamma,\beta_0}(z_{n+1:\infty}|z_{1:n}) = \nabla_{\gamma} \log p_{\gamma,\beta_0}(z_{1:n}),
\end{align*}
where $p_{\gamma,\beta_0}(z_{1:n})$ is the relevant marginal density determined from the joint density $p_{\gamma,\beta_0}(z_{1:\infty})$. For this specific sequence $z_{1:\infty},$ we therefore have
\begin{align*}
\nabla_{\gamma} \log_{\gamma,\beta_0}(z_{1:\infty}) &=  
\nabla_{\gamma} \left( \sum_{k=1}^n \log p_{\gamma,\beta_0}(z_k | z_{1:k-1}) \right) \bigg|_{\gamma=\gamma_0} \\
&= \sum_{k=1}^n \nabla_{\gamma} \log p_{\gamma,\beta_0}(z_k | z_{1:k-1}) \Big|_{\gamma=\gamma_0} \\
&= \sum_{k=1}^n \nabla_{\gamma} \log p_{\gamma,\beta_0}(z_k | z_{1:k-1}) \Big|_{\gamma=\gamma_0} + \sum_{k=n+1}^{\infty} 0 \\
&= \sum_{k=1}^{\infty} \nabla_{\gamma} \log p_{\gamma,\beta_0}(z_k | z_{1:k-1}) \Big|_{\gamma=\gamma_0}.
\end{align*}
Since the expression above holds on a subset of ${\cal X}^{(Z_{1:\infty})}$ that has probability one with respect to any of the expressions $\PR_{\gamma,\beta},$ then we can conclude that
$$h(Z_{1:\infty}) = \nabla_{\gamma} \log p_{\gamma,\beta_0}(Z_{1:\infty}) |_{\gamma=\gamma_0} = \sum_{k=1}^{\infty} S_{\gamma,k}(Z_k,\Pa(Z_k))$$
almost surely, as we claimed.

\smallskip

We now proceed to show that $BS_{\gamma,k}$ lives in ${\cal H}_k$.  Letting $\E_{\gamma,\beta}$ denote expectation with respect to $\PR_{\gamma,\beta}$, then a standard argument is that
\begin{align*}
0 = \nabla_{\gamma} 1 = \nabla_{\gamma} \E_{\gamma,\beta_0}[ 1 | \Pa(Z_k)] \Big|_{\gamma=\gamma_0}= \E_0[ S_{\gamma,k} | \Pa(Z_k)].
\end{align*}
For $k \in \mathbb{K}$, a similar argument gives
\begin{align*}
0 = \nabla_{\gamma} 0 = \nabla_{\gamma} \left( \E_{\gamma,\beta_0}[ Z_k | J_{k,1} ] - \E_{\gamma,\beta_0}[ Z_k | J_{k,0} ] \right) \Big|_{\gamma=\gamma_0} = \E_0[ S_{\gamma,k} Z_k | J_{k,1}] - \E_0[ S_{\gamma,k} Z_k | J_{k,0}].
\end{align*}
The last two expressions ensure that $BS_{\gamma,k} \in {\cal H}_k.$ We can conclude that $B h \in \oplus_{k=1}^{\infty} {\cal H}_k$ if we can argue that
$$\sum_{k=1}^{\infty} \E_0\left[ S_{\gamma,k} \cdot S_{\gamma,k} \right] < \infty.$$
Note then that $Bh \in \Gamma$ implies $Bh \in {\cal H}$ and $\E_0[h \cdot h] < \infty.$ Orthogonality of closed, linear subspaces ${\cal H}_k$ with $S_{\gamma,k} \in {\cal H}_k$ then gives
$\E_0[h \cdot h]  = \sum_{k=1}^{\infty} \E_0\left[ S_{\gamma,k} \cdot S_{\gamma, k} \right] < \infty.$
Thus, $Bh \in \oplus_{k=1}^{\infty} {\cal H}_k,$ completing our proof of the characterization of $\Gamma$. 
\EndPf

\textbf{Step 2. Characterizing the score function $S_{\beta}$.} 

We next characterize the score function $S_{\beta}$ with respect to our inferential target $\beta$. We revert back to the old labels of the variables.

\begin{applem}{(Score function $S_{\beta}$)} \label{lem:score}
    The score function $S_{\beta}$ with respect to ${\cal M}$ and the inferential target $\beta_0$ is given by
    \begin{align*}
\sum_{k=1}^{\infty} S_{\beta,k}({{Q}}_{k+\ell},\Pa({{Q}}_{k+\ell})),
\end{align*}
for functions $S_{\beta,k}$ that satisfy 
\begin{align*}
    0 &=\E_0[ S_{\beta,k} | \Pa({{Q}}_{k+\ell}) ] \\
    \mathbbm{1}_{\{T_{k+\ell}<\infty\}} f(R_k) &= \E_0[ S_{\beta,k} \, {{Q}}_{k+\ell} | A_k=1, R_k, T_{k+\ell}< \infty ] - \E_0[ S_{\beta,k} {{Q}}_{k+\ell} | A_k=0, R_k, T_{k+\ell}< \infty ].
\end{align*}
\end{applem}

We would like to comment on this lemma. In essence, $S_{\beta}$ is nearly contained within $\Gamma$. However, the moment restriction (Equation~\ref{eq:restricted_moment2}) is not completely satisfied for $S_{\beta,k}$. Instead of the left hand side of the moment restriction be zero, we observe $\mathbbm{1}_{\{T_{k+\ell}<\infty\}} f(R_k)$. To derive the efficient score function, it is necessary to correct this discrepancy.

\Pf
 We examine probability distributions $\PR_{\beta}$ within ${\cal M}$ constructed from $\beta$ and the Markov kernels $\PR_0^{(Z)}.$ Let $p_{\beta}$ denote the density and $\E_{\beta}$ the expectation with respect to a specific $\PR_{\beta}$. From the definition of ${\cal M},$ we know there exists a score function  $S_{\beta} \in {\cal H}$ such that
    $$ S_{\beta} = \nabla_{\beta} \log p_{\beta} \big|_{\beta = \beta_0}.$$
We can apply identical logic from our proof in Lemma~\ref{lem:gamma} to argue that  
\begin{align*}
   \nabla_{\beta} \log p_{\beta} \big|_{\beta = \beta_0} = \sum_{k=1}^{\infty} \nabla_{\beta} \log p_{\beta}( {{Q}}_{k+\ell} | \Pa({{Q}}_{k+\ell})) \big|_{\beta = \beta_0}, 
\end{align*}
since the sum is finite on sets that have probability one irrespective of the distribution $\PR_{\beta}$. The rest of the proof is computational in nature. 

We let
\begin{align*}
    S_{\beta,k} = S_{\beta,k}({{Q}}_{k+\ell},\Pa({{Q}}_{k+\ell})) = \nabla_{\beta} \log p_{\beta}({{Q}}_{k+\ell} | \Pa({{Q}}_{k+\ell})) \Big|_{\beta = \beta_0} = \frac{ \nabla_{\beta} \, p_{\beta}({{Q}}_{k+\ell} | \Pa({{Q}}_{k+\ell})) }{p_{\beta}({{Q}}_{k+\ell} | \Pa({{Q}}_{k+\ell}))} \bigg|_{\beta = \beta_0}
\end{align*}
for $k \in N$. Then, the same trick from the last lemma allows us to swap $\E_\beta$ for $\E_0$ in
\begin{align*}
    0 = \nabla_{\beta} 1 \big|_{\beta=\beta_0} = \nabla_{\beta} \E_{\beta}\left[ 1 | \Pa({{Q}}_{k+\ell})\right] \big|_{\beta=\beta_0} = \E_{0}\left[ S_{\beta,k} | \Pa({{Q}}_{k+\ell}) \right].
\end{align*}
A similar argument applied to the moment restriction (Equation~\ref{eq:restricted_moment2}) gives another relation:
\begin{align*}
    0 = \nabla_{\beta} 0 \big|_{\beta=\beta_0} &= \nabla_{\beta} \left( \E_{\beta}\left[ {{Q}}_{k+\ell} | A_k=1, R_k, T_{k+\ell} < \infty \right] - \E_{\beta}\left[ {{Q}}_{k+\ell} | A_k=0, R_k, T_{k+\ell} < \infty\right] \right) \big|_{\beta=\beta_0}\\
    &= \E_{\beta}\left[ \nabla_{\beta} {{Q}}_{k+\ell} \big|_{\beta=\beta_0} \Big| A_k=1, R_k, T_{k+\ell} < \infty \right] - \E_{\beta}\left[ \nabla_{\beta} {{Q}}_{k+\ell} \big|_{\beta=\beta_0} \Big| A_k=0, R_k, T_{k+\ell} < \infty\right]     \\
    &+ \E_{0}\left[ {{Q}}_{k+\ell} S_{\beta,k}| A_k=1, R_k, T_{k+\ell} < \infty \right] - \E_{0}\left[ {{Q}}_{k+\ell} S_{\beta,k} | A_k=0, R_k, T_{k+\ell} < \infty\right] \\
    &= -\mathbbm{1}_{\{T_{k+\ell}<\infty\}} f(R_k) 
    \\
    &+ \E_{0}\left[ {{Q}}_{k+\ell} S_{\beta,k}| A_k=1, R_k, T_{k+\ell} < \infty \right] - \E_{0}\left[ {{Q}}_{k+\ell} S_{\beta,k} | A_k=0, R_k, T_{k+\ell} < \infty\right].
\end{align*}
This completes the proof. 
\EndPf

\textbf{Step 3. Decompose $S_{\beta}$ into an element in $\Gamma$ and an element in $\Gamma^{\perp}$.} 

Our next lemma is also computational in nature.

\begin{applem}{(Decomposition of $S_{\beta}$)} \label{lem:dec_score}
    The score function $S_{\beta}$ with respect to ${\cal M}$ and the inferential target $\beta_0$ can be decomposed as
    \begin{align*}
(S_{\beta}-S_{\beta}^{\perp}) + S_{\beta}^{\perp},
\end{align*}
with $(S_{\beta}-S_{\beta}^{\perp}) \in \Gamma$ and $S_{\beta}^{\perp} \in \Gamma^{\perp},$
where 
\begin{align*}
    S_{\beta}^{\perp} &= \sum_{k=1}^{\infty} \frac{\mathbbm{1}_{\{T_{k+\ell}<\infty\}} f(R_k)}{\sigma_k(R_k)} (A_k-\rho_k(R_k)) {{Q}}_{k+\ell} , \\
    \rho_k(R_k) &= \E_0[A_k | R_k, T_{k+\ell}<\infty], \\
    \sigma_k(R_k) &= (1-\rho_k(R_k)) \E_0[{{Q}}_{k+\ell}^2 | A_k=1,R_k,T_{k+\ell}<\infty] + \rho_k(R_k) \E_0[{{Q}}_{k+\ell}^2 | A_k=0,R_k,T_{k+\ell}<\infty].
\end{align*}
\end{applem}

\Pf
We first want to show that $S_{\beta}-S_{\beta}^{\perp} \in \Gamma$. If we let
\begin{align*}
    S_{\beta,k}^{\perp} = \frac{\mathbbm{1}_{\{T_{k+\ell}<\infty\}} f(R_k)}{\sigma_k(R_k)}(A_k-\rho_k(R_k)) {{Q}}_{k+\ell},
\end{align*}
then because of Equation~\ref{eq:kappa_centered}, 
\begin{align*}
    &\E_0\left[ S_{\beta,k}^{\perp} | \Pa({{Q}}_{k+\ell})\right] \\
    &=     \frac{\mathbbm{1}_{\{T_{k+\ell}<\infty\}} f(R_k)}{\sigma_k(R_k)}(A_k-\rho_k(R_k)) \E_0\left[ {{Q}}_{k+\ell} | \Pa({{Q}}_{k+\ell}), T_{k+\ell} < \infty\right] \\
    &= 0.
\end{align*}
This shows that $S_{\beta,k}^{\perp}$ satisfies the first condition needed to be in $\Gamma$. Hence, $S_{\beta,k} - S_{\beta,k}^{\perp}$ also satisfies the first condition, because, according to Lemma~\ref{lem:score}, $S_{\beta,k}$ also satisfies this condition. Regarding the second condition, we have
\begin{align*}
    &\E_0\left[ S_{\beta,k}^{\perp} Q_{k+\ell} | A_k=a, R_k, T_{k+\ell} < \infty \right] \\
    &=\frac{\mathbbm{1}_{\{T_{k+\ell}<\infty\}} f(R_k)}{\sigma_k(R_k)} 
    (a-\rho_k(R_k))\E_0\left[{{Q}}_{k+\ell}^2 | A_k=a, R_k, T_{k+\ell} < \infty \right].
\end{align*}
Therefore, 
\begin{align*}
    &\E_0\left[ S_{\beta,k}^{\perp} Q_{k+\ell} | A_k=1, R_k, T_{k+\ell} < \infty \right] - \E_0\left[ S_{\beta,k}^{\perp} Q_{k+\ell} | A_k=0, R_k, T_{k+\ell} < \infty \right] = \mathbbm{1}_{\{T_{k+\ell}<\infty\}} f(R_k),  
\end{align*}
which underscores why we defined $\sigma_k(R_k)$ as we did. Combining this expression with Lemma~\ref{lem:score}, we get
\begin{align*}
    &\E_0\left[ S_{\beta,k}-S_{\beta,k}^{\perp} | A_k=1, R_k, T_{k+\ell} < \infty \right] - \E_0\left[ S_{\beta,k}-S_{\beta,k}^{\perp} | A_k=0, R_k, T_{k+\ell} < \infty \right] \\
    &= \mathbbm{1}_{\{T_{k+\ell}<\infty\}} f(R_k) - \mathbbm{1}_{\{T_{k+\ell}<\infty\}} f(R_k) = 0.  
\end{align*}
We conclude that $S_{\beta,k}-S_{\beta,k}^{\perp} \in \Gamma$, and also, because $\Gamma$ is closed and linear, that $\sum_{k=1}^{\infty}(S_{\beta,k}-S_{\beta,k}^{\perp})=S_{\beta}-S_{\beta}^{\perp} \in \Gamma$.

We can finish the proof, provided we can show that $S_{\beta}^{\perp} \in \Gamma^{\perp}.$  It suffices to show that $S_{\beta,k}^{\perp} \in \Gamma^{\perp}$, because $\Gamma^{\perp}$ is closed and linear. For this last step, we lean on our relabeling of variables (the $Z_i$) from our characterization of $\Gamma$ (Lemma~\ref{lem:gamma}), in which case take any $h_j(Z_j,\Pa(Z_j)) \in {\cal H}_j \subseteq \Gamma.$ Our relabeling will also allow us to write $S_{\beta,k}^{\perp}$ as a function of the form $g_{i}(Z_{i},\Pa(Z_{i}))$ for the appropriate $i \in \mathbb{K}$. From our work above, we know
$$\E_0[S_{\beta,k}^{\perp} | \Pa({{Q}}_{k+\ell})] =\E_0[ h_{i} | \Pa(Z_i) ] = 0.$$
If $i < j$, then using an argument from our characterization of $\Gamma$ (Lemma~\ref{lem:gamma}), we have
\begin{align*}
    \E_0\left[ h_j \cdot g_i \right] = \E_0\left[ \E_0[h_j \cdot g_i | Z_{1:j-1}] \right] = \E_0\left[ \E_0[h_j | \Pa(Z_j)] \cdot g_i \right] = 0.
\end{align*}
If $j < i$, the same logic implies
\begin{align*}
    \E_0\left[ h_j \cdot g_i \right] = 0.
\end{align*}
The final case to consider is $i=j$, in which case
\begin{align*}
    & \E_0\left[ h_i \cdot g_i | A_k=a,  R_k, T_{k+\ell}<\infty \right] \\
   &= \frac{\mathbbm{1}_{\{T_{k+\ell}<\infty\}}}{\sigma_k(R_k)}(a-\rho_k(R_k))  f(R_k) \cdot \E_0\left[ h_i {{Q}}_{k+\ell} | A_k=a,  R_k, T_{k+\ell}<\infty \right]
\end{align*}
and hence
\begin{align*}
    & \E_0\left[ h_i \cdot g_i |  R_k, T_{k+\ell}<\infty \right] = \frac{\mathbbm{1}_{\{T_{k+\ell}<\infty\}}}{\sigma_k(R_k)}(1-\rho_k(R_k))\rho_k(R_k) \times \\
   & \qquad \qquad f(R_k) \cdot \left( \E_0\left[ h_i {{Q}}_{k+\ell} | A_k=1,  R_k, T_{k+\ell}<\infty \right] - \E_0\left[ h_i {{Q}}_{k+\ell} | A_k=0,  R_k, T_{k+\ell}<\infty \right]\right).
\end{align*}
Notice that upon relabeling, 
\begin{align*}
    &\E_0\left[ h_i {{Q}}_{k+\ell} | A_k=1,  R_k, T_{k+\ell}<\infty \right] - \E_0\left[ h_i {{Q}}_{k+\ell} | A_k=0,  R_k, T_{k+\ell}<\infty \right] \\
    &= \E_0\left[ h_i Z_i | J_{i,1} \right] - \E_0\left[ h_i Z_i | J_{i,0} \right],
\end{align*}
which is equal to 0, because of what it means for $h_i \in {\cal H}_i$. This allows us to arrive at
\begin{align*}
    &\E_0\left[ h_i \cdot g_i \right] =\E_0\left[ \E_0\left[ h_i \cdot g_i | R_k, T_{k+\ell} < \infty \right]\right] = 0,
\end{align*}
demonstrating that $S_{\beta,k}=g_i$ is orthogonal to arbitrary $h_j \in {\cal H}_j$ and arbitrary $j$. Any $h \in \Gamma$ is an (infinite) sum of such $h_j$ or limit points of such sums, which means $S_{\beta,k}^{\perp}$ is orthogonal to arbitrary $h \in \Gamma$. By definition of orthogonal complement, $S_{\beta,k}^{\perp} \in \Gamma^{\perp}$. As $\Gamma^{\perp}$ is closed in ${\cal H}$,  $S_{\beta}^{\perp} = \sum_{k=1}^{\infty} S_{\beta,k}^{\perp} \in \Gamma^{\perp}$. This completes this lemma's proof.
\EndPf

\textbf{Step 4. Putting everything together to prove theorem.} 

\textbf{Proof of Theorem~\mref{thm:opt_efficiency}.}

Our proof is now nearly complete. We have obtained the nuisance tangent space $\Gamma$ from Lemma~\ref{lem:gamma} and the score function $S_{\beta}$ from Lemma~\ref{lem:score}. The semiparametric efficiency score $S_{\rm{eff}}(\beta)$ is defined as $S_{\beta} - \Pi(S_{\beta} | \Gamma)$, where $\Pi(S_{\beta} | \Gamma)$ represents the orthogonal projection of $S_{\beta}$ onto $\Gamma$. In Lemma~\ref{lem:dec_score}, we show that $S_{\beta}$ can be decomposed into an element $(S_{\beta}-S_{\beta}^{\perp})$ within $\Gamma$ and an element $S_{\beta}^{\perp}$ within $\Gamma^{\perp}$. Since $\Gamma$ is a closed linear subspace of ${\cal H}$, this decomposition is unique, and the element $S_{\beta}-S_{\beta}^{\perp}$ must be equal to $\Pi(S_{\beta} | \Gamma)$. Thus, we have $S_{\rm{eff}}(\beta) = S_{\beta}^{\perp}$.

To finalize the proof, we need to express $S_{\beta}^{\perp}$ in the form given in the theorem. For this, consider
\begin{align*}
    \mu_k(R_k) &= \E_0[Y_k | A_k=0, R_k, T_{k+\ell} < \infty] \\
    \rho_k(R_k) &= \E_0[A_k | R_k, T_{k+\ell}<\infty], \\
    \sigma_k(R_k) &= (1-\rho_k(R_k)) \E_0[{{Q}}_{k+\ell}^2 | A_k=1,R_k,T_{k+\ell}<\infty] + \rho_k(R_k) \E_0[{{Q}}_{k+\ell}^2 | A_k=0,R_k,T_{k+\ell}<\infty] \\
    K &= \sup \{ k \in N : T_{k} < \infty \}.
\end{align*}
We can rewrite $\sigma_k(R_k)$ as
\begin{align*}
    \sigma_k(R_k) &= (1-\rho_k(R_k)) {\rm{Var}_0}[Y_{k+\ell} | A_k=1,R_k,T_{k+\ell}<\infty] + \rho_k(R_k) {\rm{Var}_0}[Y_{k+\ell} | A_k=0,R_k,T_{k+\ell}<\infty]
\end{align*}
from the definition of ${{Q}}_{k+\ell}$ and Equation~\ref{eq:kappa_centered}, where variance is taken with respect to $\PR_0.$ This leads us to:
\begin{align*}
    S_{\rm{eff}}(\beta) = S_{\beta}^{\perp} &= \sum_{k=1}^{\infty} \frac{\mathbbm{1}_{\{T_{k+\ell}<\infty\}} f(R_k)}{\sigma_k(R_k)} (A_k-\rho_k(R_k)) {{Q}}_{k+\ell} \\
    &= \sum_{k=1}^{K-\ell} \frac{f(R_k)}{\sigma_k(R_k)} (A_k-\rho_k(R_k)) (Y_{k+\ell} - \mu_k(R_k) - A_k f(R_k)'\beta).
\end{align*}
This expression precisely matches the target form stated in the theorem.
\EndPf

\section{Geex Implementation}\label{app:geex}

This section presents the code used to calculate the lag-1 effect of directing a patient to a vertical area on the subsequent patient's outcome. This estimation was performed using the geex package \citep{geex}, which offers a convenient application programming interface (API) for conducting Z-estimation. The API requires the analyst to provide a specific function known as \emph{estFun}, which maps the data to an estimating function representing the parameter of interest. The subsequent code snippet represents the implementation of the required \emph{estFun} function in the R programming language.

\begin{lstlisting}[language=R]
# This R function represents the estimating equation 
# E_n [U_stacked] = 0.

# data: A single cluster/panel of data. 
# Data must contain the intervention column (is_vertical)
# and the covariates used to fit the nuisance models

# models: A list of fitted nuisance models:
#    * mlagged = fitted model for the future outcome,  
#      E[Y_{k+\ell}|R_k] = g(R_k)'\alpha (includes f(S_k)'\beta')
#    * mpk = fitted model for denominators in weights,
#      E[A_k|R_k, T_{k+\ell} < \infty] = pk(R_k; \eta)
#    * mqk = fitted model for numerators in weights,
#      E[A_k|S_k, T_{k+\ell} < \infty]= qk(S_k; \xi)

# M: The number of independent clusters

# This function returns a function f(data, \theta), with
# theta = (\alpha, \beta, \xi, \eta)
# such that f(data,\theta) = U_stacked(data, \theta)

est_fun <- function(data, models, M){    
    # model for future outcome
    mlagged <- models$mlagged

    # model for denominators of weights
    mpk <- models$mpk

    # model for numerator of weights
    mqk <- models$mqk
    
    # extracting variables used in estimating equation (K = cluster size)
    Ak <- data$is_vertical   # (K-1) x 1
    
    # extracting outcome name and values
    formula <- mlagged$formula
    outcome <- all.vars(formula)[1]
    Ylagged <- data[,c(outcome)]   # (K-1) x 1
    
    # design matrices of nuisance models
    
    # (K-1) x (alpha + beta)
    Xlagged <- grab_design_matrix(data = data, rhs_formula = grab_fixed_formula(mlagged)) 
    # (K-1) x xi
    Xqk <- grab_design_matrix(data = data, rhs_formula = grab_fixed_formula(mqk))           
    # (K-1) x eta
    Xpk <- grab_design_matrix(data = data, rhs_formula = grab_fixed_formula(mpk))            
    # theta = (alpha, beta, xi, eta)
    alpha_beta_pos <- 1:ncol(Xlagged)                                           
    xi_pos <- (max(alpha_beta_pos)+1):(max(alpha_beta_pos) + ncol(Xqk))         
    eta_pos <- (max(xi_pos)+1):(max(xi_pos) + ncol(Xpk))                       
    
    # Estimating functions for logistic models
    qk_scores <- grab_psiFUN(mqk, data)
    pk_scores <- grab_psiFUN(mpk, data)
    
    function (theta){
        # fitted values of the models g(R_k)'alpha, q_k and p_k respectively
        lagged <- Xlagged %*% theta[alpha_beta_pos]  # (K-1) x 1
        qk <- plogis(Xqk %*% theta[xi_pos])          # (K-1) x 1
        pk <- plogis(Xpk %*% theta[eta_pos])         # (K-1) x 1
        
        # Weights
        Wk <- Ak*(qk/pk) + (1-Ak)*((1-qk)/(1-pk))   # (K-1) x 1 
        
        # A = Wk(Y_{k+\ell} - g(Rk)'alpha - Ak f(Sk)'beta)
        A <- Wk*(Ylagged - lagged)  # (K-1) x 1
        
        repA <- matrix(replicate(ncol(Xlagged),A), nrow=length(A)) 

        #  A * [g(R_k), A_k f(S_k)]' of size (K-1) x (alpha + beta) 
        Uk <- repA * Xlagged 
        
        # Adding the contribution of each individual job
        SumUk <- colSums(Uk)
        
        # (alpha + beta) + xi + eta
        c(SumUk/M, 
          qk_scores(theta[xi_pos])/M, 
          pk_scores(theta[eta_pos])/M) 
    }
}
\end{lstlisting}

Once the necessary \textit{estFun} function has been provided, we can estimate standard errors. We assume that the nuisance models for the future outcome (\emph{lagged}), as well as the numerators (\emph{mqk}) and the denominators (\emph{mpk}) of weights, have been appropriately fitted. We can invoke the geex package, which facilitates this procedure:

\begin{lstlisting}[language=R]
# Number of independent clusters
M <- max(data$arrival_day_order)

# fitted coefficients
alpha_beta <- coef(mlagged)
xi <- coef(mqk)
eta <- coef(mpk)
                         
theta <- c(alpha_beta, xi, eta)

# list of fitted nuisance models
models <- list(mpk=mpk, mqk=mqk, mlagged=mlagged)

mlag <- m_estimate(estFUN = est_fun,
                   data = data,
                   units = 'arrival_day_order',
                   roots = theta,
                   compute_roots = FALSE,
                   outer_args = list(models = models, M=M))

# covariance matrix
vcov(mlag)
\end{lstlisting}

%% file: references.bib
@article{33454a05-596e-3e2d-a725-b726efc21380,
 ISSN = {13507265},
 author = {Kjetil R{\o}ysland},
 journal = {Bernoulli},
 number = {3},
 pages = {895--915},
 publisher = {International Statistical Institute (ISI) and Bernoulli Society for Mathematical Statistics and Probability},
 title = {A martingale approach to continuous-time marginal structural models},
 urldate = {2023-06-12},
 volume = {17},
 year = {2011}
}

@misc{RTC-Dec2018,
title = {2019 {R}isk {A}djustment {F}actors and {P}ayment {R}ates},
howpublished = {\url{https://www.cms.gov/Medicare/Health-Plans/MedicareAdvtgSpecRateStats/Downloads/RTC-Dec2018.pdf}},
year = {2018},
note = {Accessed: March 23, 2023}
}

@article{alagoz2010markov,
  title={Markov decision processes: a tool for sequential decision making under uncertainty},
  author={Alagoz, Oguzhan and Hsu, Heather and Schaefer, Andrew J and Roberts, Mark S},
  journal={Medical Decision Making},
  volume={30},
  number={4},
  pages={474--483},
  year={2010},
  publisher={SAGE Publications Sage CA: Los Angeles, CA}
}

@article{aronow_estimating_2017,
	title = {Estimating average causal effects under general interference, with application to a social network experiment},
	volume = {11},
	issn = {1932-6157},
	doi = {10.1214/16-AOAS1005},
	number = {4},
	journal = {The Annals of Applied Statistics},
	author = {Aronow, Peter M. and Samii, Cyrus},
	month = dec,
	year = {2017},
}

@article{boruvka_assessing_2018,
	title = {Assessing {Time}-{Varying} {Causal} {Effect} {Moderation} in {Mobile} {Health}},
	volume = {113},
	issn = {0162-1459, 1537-274X},
	doi = {10.1080/01621459.2017.1305274},
	language = {en},
	number = {523},
	journal = {Journal of the American Statistical Association},
	author = {Boruvka, Audrey and Almirall, Daniel and Witkiewitz, Katie and Murphy, Susan A.},
	month = jul,
	year = {2018},
	pages = {1112--1121},
}

@article{chakraborty2013statistical,
  title={Statistical methods for dynamic treatment regimes},
  author={Chakraborty, Bibhas and Moodie, Erica E},
  journal={Springer-Verlag. doi},
  volume={10},
  pages={978--1},
  year={2013},
  publisher={Springer}
}

@article{chakraborty_dynamic_2014,
	title = {Dynamic {Treatment} {Regimes}},
	volume = {1},
	issn = {2326-8298, 2326-831X},
	doi = {10.1146/annurev-statistics-022513-115553},
	language = {en},
	number = {1},
	journal = {Annual Review of Statistics and Its Application},
	author = {Chakraborty, Bibhas and Murphy, Susan A.},
	month = jan,
	year = {2014},
	pages = {447--464},
}

@article{chen_decision_2016,
	title = {Decision {Making} {Under} the {Gambler}’s {Fallacy}: {Evidence} from {Asylum} {Judges}, {Loan} {Officers}, and {Baseball} {Umpires}*},
	volume = {131},
	issn = {0033-5533, 1531-4650},
	shorttitle = {Decision {Making} {Under} the {Gambler}’s {Fallacy}},
	doi = {10.1093/qje/qjw017},
	language = {en},
	number = {3},
	journal = {The Quarterly Journal of Economics},
	author = {Chen, Daniel L. and Moskowitz, Tobias J. and Shue, Kelly},
	month = aug,
	year = {2016},
	pages = {1181--1242},
}

@article{cochran2023mobile,
  title={Mobile Acceptance and Commitment Therapy in Bipolar Disorder: Microrandomized Trial},
  author={Cochran, Amy and Maronge, Jacob M and Victory, Amanda and Hoel, Sydney and McInnis, Melvin G and Thomas, Emily BK and others},
  journal={JMIR Mental Health},
  volume={10},
  number={1},
  pages={e43164},
  year={2023},
  publisher={JMIR Publications Inc., Toronto, Canada}
}

@book{cox_planning_1958,
	address = {Oxford, England},
	series = {Planning of experiments},
	title = {Planning of experiments},
	publisher = {Wiley},
	author = {Cox, D. R.},
	year = {1958},
	note = {Pages: 308},
}

@article{didelez2008graphical,
  title={Graphical models for marked point processes based on local independence},
  author={Didelez, Vanessa},
  journal={Journal of the Royal Statistical Society: Series B (Statistical Methodology)},
  volume={70},
  number={1},
  pages={245--264},
  year={2008},
  publisher={Wiley Online Library}
}

@inproceedings{didelez2015causal,
  title={Causal Reasoning for Events in Continuous Time: A Decision-Theoretic Approach.},
  author={Didelez, Vanessa},
  booktitle={ACI@ UAI},
  pages={40--45},
  year={2015}
}

@article{gao2021causal,
  title={Causal inference for event pairs in multivariate point processes},
  author={Gao, Tian and Subramanian, Dharmashankar and Bhattacharjya, Debarun and Shou, Xiao and Mattei, Nicholas and Bennett, Kristin P},
  journal={Advances in Neural Information Processing Systems},
  volume={34},
  pages={17311--17324},
  year={2021}
}

@article{garrett2018effect,
  title={The effect of vertical split-flow patient management on emergency department throughput and efficiency},
  author={Garrett, John S and Berry, Colyn and Wong, Hao and Qin, Huanying and Kline, Jeffery A},
  journal={The American Journal of Emergency Medicine},
  year={2018},
  publisher={Elsevier}
}

@article{geex,
    title = {The Calculus of M-Estimation in {R} with {geex}},
    author = {Bradley C. Saul and Michael G. Hudgens},
    journal = {Journal of Statistical Software},
    year = {2020},
    volume = {92},
    number = {2},
    pages = {1--15},
    doi = {10.18637/jss.v092.i02},
  }

@article{gill_causal_2001,
	title = {Causal {Inference} for {Complex} {Longitudinal} {Data}: {The} {Continuous} {Case}},
	volume = {29},
	issn = {0090-5364},
	shorttitle = {Causal {Inference} for {Complex} {Longitudinal} {Data}},
	number = {6},
	urldate = {2022-11-11},
	journal = {The Annals of Statistics},
	author = {Gill, Richard D. and Robins, James M.},
	year = {2001},
	note = {Publisher: Institute of Mathematical Statistics},
	pages = {1785--1811},
}

@article{goldbach_sequential_2022,
	title = {Sequential decision bias – evidence from grading exams},
	volume = {54},
	issn = {0003-6846, 1466-4283},
	doi = {10.1080/00036846.2021.1976390},
	language = {en},
	number = {32},
	journal = {Applied Economics},
	author = {Goldbach, Carina and Sickmann, Jörn and Pitz, Thomas},
	month = jul,
	year = {2022},
	pages = {3727--3739},
}

@article{gomez2022evaluation,
  title={Evaluation of a Split Flow Model for the Emergency Department},
  author={David, Juan Camilo and Cochran, Amy L and Patterson, Brian W and Zayas-Caban, Gabriel},
  journal={arXiv preprint arXiv:2202.00736},
  year={2022}
}

@article{guo2021discussion,
  title={Discussion of ‘Estimating time-varying causal excursion effects in mobile health with binary outcomes’},
  author={Guo, F Richard and Richardson, Thomas S and Robins, James M},
  journal={Biometrika},
  volume={108},
  number={3},
  pages={541--550},
  year={2021},
  publisher={Oxford University Press}
}

@book{hinderer2016dynamic,
  title={Dynamic optimization},
  author={Hinderer, Karl and Rieder, Ulrich and Stieglitz, Michael},
  year={2016},
  publisher={Springer}
}

@article{hong_evaluating_2006,
	title = {Evaluating {Kindergarten} {Retention} {Policy}: {A} {Case} {Study} of {Causal} {Inference} for {Multilevel} {Observational} {Data}},
	volume = {101},
	issn = {0162-1459, 1537-274X},
	shorttitle = {Evaluating {Kindergarten} {Retention} {Policy}},
	doi = {10.1198/016214506000000447},
	language = {en},
	number = {475},
	journal = {Journal of the American Statistical Association},
	author = {Hong, Guanglei and Raudenbush, Stephen W},
	month = sep,
	year = {2006},
	pages = {901--910},
}

@article{horvitz1952generalization,
  title={A generalization of sampling without replacement from a finite universe},
  author={Horvitz, Daniel G and Thompson, Donovan J},
  journal={Journal of The American Statistical Association},
  volume={47},
  number={260},
  pages={663--685},
  year={1952},
  publisher={Taylor \& Francis}
}

@article{hudgens_toward_2008,
	title = {Toward {Causal} {Inference} {With} {Interference}},
	volume = {103},
	issn = {0162-1459, 1537-274X},
	doi = {10.1198/016214508000000292},
	language = {en},
	number = {482},
	journal = {Journal of the American Statistical Association},
	author = {Hudgens, Michael G and Halloran, M. Elizabeth},
	month = jun,
	year = {2008},
	pages = {832--842},
}

@article{jacobsen2006point,
  title={Point process theory and applications: marked point and piecewise deterministic processes},
  author={Jacobsen, Martin and Gani, Joseph},
  year={2006},
  publisher={Springer}
}

@article{klasnja_micro-randomized_2015,
	title = {Micro-{Randomized} {Trials}: {An} {Experimental} {Design} for {Developing} {Just}-in-{Time} {Adaptive} {Interventions}},
	volume = {34},
	issn = {0278-6133},
	shorttitle = {Micro-{Randomized} {Trials}},
	doi = {10.1037/hea0000305},
	number = {0},
	urldate = {2022-02-17},
	journal = {Health Psychology : Official Journal of the Division of Health Psychology, American Psychological Association},
	author = {Klasnja, Predrag and Hekler, Eric B. and Shiffman, Saul and Boruvka, Audrey and Almirall, Daniel and Tewari, Ambuj and Murphy, Susan A.},
	month = dec,
	year = {2015},
	pmid = {26651463},
	pmcid = {PMC4732571},
	pages = {1220--1228},
}

@article{konrad2013modeling,
  title={Modeling the impact of changing patient flow processes in an emergency department: Insights from a computer simulation study},
  author={Konrad, Renata and DeSotto, Kristine and Grocela, Allison and McAuley, Patrick and Wang, Justin and Lyons, Jill and Bruin, Michael},
  journal={Operations Research for Health Care},
  volume={2},
  number={4},
  pages={66--74},
  year={2013},
  publisher={Elsevier}
}

@article{laber2014dynamic,
  title={Dynamic treatment regimes: Technical challenges and applications},
  author={Laber, Eric B and Lizotte, Daniel J and Qian, Min and Pelham, William E and Murphy, Susan A},
  journal={Electronic Journal of Statistics},
  volume={8},
  number={1},
  pages={1225},
  year={2014},
  publisher={NIH Public Access}
}

@article{lavori2004dynamic,
  title={Dynamic treatment regimes: practical design considerations},
  author={Lavori, Philip W and Dawson, Ree},
  journal={Clinical trials},
  volume={1},
  number={1},
  pages={9--20},
  year={2004},
  publisher={Sage Publications Sage CA: Thousand Oaks, CA}
}

@article{lee2011weight,
  title={Weight trimming and propensity score weighting},
  author={Lee, Brian K and Lessler, Justin and Stuart, Elizabeth A},
  journal={PloS one},
  volume={6},
  number={3},
  pages={e18174},
  year={2011},
  publisher={Public Library of Science San Francisco, USA}
}

@article{li2019addressing,
  title={Addressing extreme propensity scores via the overlap weights},
  author={Li, Fan and Thomas, Laine E and Li, Fan},
  journal={American Journal of Epidemiology},
  volume={188},
  number={1},
  pages={250--257},
  year={2019},
  publisher={Oxford University Press}
}

@article{li2023optimal,
  title={Optimal treatment regimes: a review and empirical comparison},
  author={Li, Zhen and Chen, Jie and Laber, Eric and Liu, Fang and Baumgartner, Richard},
  journal={International Statistical Review},
  year={2023},
  publisher={Wiley Online Library}
}

@inproceedings{malinsky2019potential,
  title={A potential outcomes calculus for identifying conditional path-specific effects},
  author={Malinsky, Daniel and Shpitser, Ilya and Richardson, Thomas},
  booktitle={The 22nd International Conference on Artificial Intelligence and Statistics},
  pages={3080--3088},
  year={2019},
  organization={PMLR}
}

@article{murphy2005experimental,
  author = {Murphy, Stephen A.},
  title = {An experimental design for the development of adaptive treatment strategies},
  journal = {Statistics in Medicine},
  volume = {24},
  number = {19},
  pages = {3179--3194},
  year = {2005},
  doi = {10.1002/sim.2022},
}

@article{nahum-shani_just--time_2017,
	title = {Just-in-{Time} {Adaptive} {Interventions} ({JITAIs}) in {Mobile} {Health}: {Key} {Components} and {Design} {Principles} for {Ongoing} {Health} {Behavior} {Support}},
	volume = {52},
	issn = {0883-6612},
	shorttitle = {Just-in-{Time} {Adaptive} {Interventions} ({JITAIs}) in {Mobile} {Health}},
	doi = {10.1007/s12160-016-9830-8},
	number = {6},
	urldate = {2022-02-18},
	journal = {Annals of Behavioral Medicine: A Publication of the Society of Behavioral Medicine},
	author = {Nahum-Shani, Inbal and Smith, Shawna N and Spring, Bonnie J and Collins, Linda M and Witkiewitz, Katie and Tewari, Ambuj and Murphy, Susan A},
	month = dec,
	year = {2017},
	pmid = {27663578},
	pmcid = {PMC5364076},
	pages = {446--462},
}

@article{neyman1923application,
  title={On the application of probability theory to agricultural experiments. Essay on principles},
  author={Neyman, Jerzy},
  journal={Ann. Agricultural Sciences},
  pages={1--51},
  year={1923}
}

@article{pan2001akaike,
  title={Akaike's information criterion in generalized estimating equations},
  author={Pan, Wei},
  journal={Biometrics},
  volume={57},
  number={1},
  pages={120--125},
  year={2001},
  publisher={Wiley Online Library}
}

@article{pearl2015conditioning,
  title={Conditioning on post-treatment variables},
  author={Pearl, Judea},
  journal={Journal of Causal Inference},
  volume={3},
  number={1},
  pages={131--137},
  year={2015},
  publisher={De Gruyter}
}

@book{pearl_causality_2000,
	address = {Cambridge, U.K. ; New York},
	title = {Causality: models, reasoning, and inference},
	isbn = {978-0-521-89560-6 978-0-521-77362-1},
	shorttitle = {Causality},
	publisher = {Cambridge University Press},
	author = {Pearl, Judea},
	year = {2000},
}

@article{philip2000,
 ISSN = {09641998, 1467985X},
 author = {Philip W. Lavori and Ree Dawson},
 journal = {Journal of the Royal Statistical Society. Series A (Statistics in Society)},
 number = {1},
 pages = {29--38},
 publisher = {[Wiley, Royal Statistical Society]},
 title = {A Design for Testing Clinical Strategies: Biased Adaptive within-Subject Randomization},
 volume = {163},
 year = {2000}
}

@article{qian_estimating_2021,
	title = {Estimating time-varying causal excursion effects in mobile health with binary outcomes},
	volume = {108},
	issn = {0006-3444, 1464-3510},
	doi = {10.1093/biomet/asaa070},
	language = {en},
	number = {3},
	urldate = {2022-11-11},
	journal = {Biometrika},
	author = {Qian, Tianchen and Yoo, Hyesun and Klasnja, Predrag and Almirall, Daniel and Murphy, Susan A},
	month = aug,
	year = {2021},
	pages = {507--527},
}

@article{richardson2013single,
  title={Single world intervention graphs (SWIGs): A unification of the counterfactual and graphical approaches to causality},
  author={Richardson, Thomas S and Robins, James M},
  journal={Center for the Statistics and the Social Sciences, University of Washington Series. Working Paper},
  volume={128},
  number={30},
  pages={2013},
  year={2013},
  publisher={Citeseer}
}

@article{robins1986new,
  title={A new approach to causal inference in mortality studies with a sustained exposure period—application to control of the healthy worker survivor effect},
  author={Robins, James},
  journal={Mathematical modelling},
  volume={7},
  number={9-12},
  pages={1393--1512},
  year={1986},
  publisher={Elsevier}
}

@inproceedings{robins_causal_1997,
	address = {New York, NY},
	series = {Lecture {Notes} in {Statistics}},
	title = {Causal {Inference} from {Complex} {Longitudinal} {Data}},
	isbn = {978-1-4612-1842-5},
	doi = {10.1007/978-1-4612-1842-5_4},
	language = {en},
	booktitle = {Latent {Variable} {Modeling} and {Applications} to {Causality}},
	publisher = {Springer},
	author = {Robins, James M.},
	editor = {Berkane, Maia},
	year = {1997},
	pages = {69--117},
}

@article{robins_correcting_1994,
	title = {Correcting for non-compliance in randomized trials using structural nested mean models},
	volume = {23},
	issn = {0361-0926, 1532-415X},
	doi = {10.1080/03610929408831393},
	language = {en},
	number = {8},
	journal = {Communications in Statistics - Theory and Methods},
	author = {Robins, James M.},
	month = jan,
	year = {1994},
	pages = {2379--2412},
}

@article{roysland2012counterfactual,
  title={Counterfactual analyses with graphical models based on local independence},
  author={Kjetil R{\o}ysland},
  journal={The Annals of Statistics},
  pages={2162--2194},
  year={2012},
  publisher={JSTOR}
}

@article{rubin1974estimating,
  title={Estimating causal effects of treatments in randomized and nonrandomized studies.},
  author={Rubin, Donald B},
  journal={Journal of Educational Psychology},
  volume={66},
  number={5},
  pages={688},
  year={1974},
  publisher={American Psychological Association}
}

@article{rubin1980randomization,
  title={Randomization analysis of experimental data: The Fisher randomization test comment},
  author={Rubin, Donald B},
  journal={Journal of the American Statistical Association},
  volume={75},
  number={371},
  pages={591--593},
  year={1980},
  publisher={JSTOR}
}

@article{rubin_causal_2005,
	title = {Causal {Inference} {Using} {Potential} {Outcomes}},
	volume = {100},
	issn = {0162-1459},
	doi = {10.1198/016214504000001880},
	number = {469},
	urldate = {2022-02-16},
	journal = {Journal of the American Statistical Association},
	author = {Rubin, Donald B},
	month = mar,
	year = {2005},
	note = {Publisher: Taylor \& Francis
\_eprint: https://doi.org/10.1198/016214504000001880},
	pages = {322--331},
}

@article{shi2022estimating,
  title={Estimating Time-Varying Direct and Indirect Causal Excursion Effects with Longitudinal Binary Outcomes},
  author={Shi, Jieru and Wu, Zhenke and Dempsey, Walter},
  journal={arXiv preprint arXiv:2212.01472},
  year={2022}
}

@article{sobel_what_2006,
	title = {What {Do} {Randomized} {Studies} of {Housing} {Mobility} {Demonstrate}?: {Causal} {Inference} in the {Face} of {Interference}},
	volume = {101},
	issn = {0162-1459, 1537-274X},
	shorttitle = {What {Do} {Randomized} {Studies} of {Housing} {Mobility} {Demonstrate}?},
	doi = {10.1198/016214506000000636},
	language = {en},
	number = {476},
	journal = {Journal of the American Statistical Association},
	author = {Sobel, Michael E},
	month = dec,
	year = {2006},
	pages = {1398--1407},
}

@article{sofrygin_semi-parametric_2017,
	title = {Semi-{Parametric} {Estimation} and {Inference} for the {Mean} {Outcome} of the {Single} {Time}-{Point} {Intervention} in a {Causally} {Connected} {Population}},
	volume = {5},
	issn = {2193-3685, 2193-3677},
	doi = {10.1515/jci-2016-0003},
	language = {en},
	number = {1},
	journal = {Journal of Causal Inference},
	author = {Sofrygin, Oleg and van der Laan, Mark J.},
	month = sep,
	year = {2017},
	pages = {20160003},
}

@article{steimle2017markov,
  title={Markov decision processes for screening and treatment of chronic diseases},
  author={Steimle, Lauren N and Denton, Brian T},
  journal={Markov Decision Processes in Practice},
  pages={189--222},
  year={2017},
  publisher={Springer}
}

@article{susan2003,
 ISSN = {13697412, 14679868},
 author = {S. A. Murphy},
 journal = {Journal of the Royal Statistical Society. Series B (Statistical Methodology)},
 number = {2},
 pages = {331--366},
 publisher = {[Royal Statistical Society, Wiley]},
 title = {Optimal Dynamic Treatment Regimes},
 volume = {65},
 year = {2003}
}

@article{tauchen1985diagnostic,
  title={Diagnostic testing and evaluation of maximum likelihood models},
  author={Tauchen, George},
  journal={Journal of Econometrics},
  volume={30},
  number={1-2},
  pages={415--443},
  year={1985},
  publisher={Elsevier}
}

@article{tchetgen_tchetgen_auto-g-computation_2021,
	title = {Auto-{G}-{Computation} of {Causal} {Effects} on a {Network}},
	volume = {116},
	issn = {0162-1459, 1537-274X},
	doi = {10.1080/01621459.2020.1811098},
	language = {en},
	number = {534},
	journal = {Journal of the American Statistical Association},
	author = {Tchetgen Tchetgen, Eric J. and Fulcher, Isabel R. and Shpitser, Ilya},
	month = apr,
	year = {2021},
	pages = {833--844},
}

@article{thomas2023mobile,
  title={Mobile Acceptance and Commitment Therapy With Distressed First-Generation College Students: Microrandomized Trial},
  author={Thomas, Emily Brenny Kroska and Sagorac Gruichich, Tijana and Maronge, Jacob M and Hoel, Sydney and Victory, Amanda and Stowe, Zachary N and Cochran, Amy},
  journal={JMIR Mental Health},
  volume={10},
  pages={e43065},
  year={2023},
  publisher={JMIR Publications Toronto, Canada}
}

@book{tsiatis2019dynamic,
  title={Dynamic treatment regimes: Statistical methods for precision medicine},
  author={Tsiatis, Anastasios A and Davidian, Marie and Holloway, Shannon T and Laber, Eric B},
  year={2019},
  publisher={CRC press}
}

@book{tsiatis_semiparametric_2006,
	address = {New York},
	series = {Springer series in statistics},
	title = {Semiparametric theory and missing data},
	isbn = {978-0-387-32448-7},
	publisher = {Springer},
	author = {Tsiatis, Anastasios A.},
	year = {2006},
}

@book{van2000asymptotic,
  title={Asymptotic statistics},
  author={Van der Vaart, Aad W},
  volume={3},
  year={2000},
  publisher={Cambridge university press}
}

@article{vansteelandt2016revisiting,
  title={Revisiting g-estimation of the effect of a time-varying exposure subject to time-varying confounding},
  author={Vansteelandt, Stijn and Sjolander, Arvid},
  journal={Epidemiologic Methods},
  volume={5},
  number={1},
  pages={37--56},
  year={2016},
  publisher={De Gruyter}
}

@article{verbitsky-savitz_causal_2012,
	title = {Causal {Inference} {Under} {Interference} in {Spatial} {Settings}: {A} {Case} {Study} {Evaluating} {Community} {Policing} {Program} in {Chicago}},
	volume = {1},
	issn = {2161-962X},
	shorttitle = {Causal {Inference} {Under} {Interference} in {Spatial} {Settings}},
	doi = {10.1515/2161-962X.1020},
	number = {1},
	journal = {Epidemiologic Methods},
	author = {Verbitsky-Savitz, Natalya and Raudenbush, Stephen W.},
	month = jan,
	year = {2012},
}

@article{wang_new_2020,
	title = {New \${G}\$-formula for the sequential causal effect and blip effect of treatment in sequential causal inference},
	volume = {48},
	issn = {0090-5364},
	doi = {10.1214/18-AOS1795},
	number = {1},
	journal = {The Annals of Statistics},
	author = {Wang, Xiaoqin and Yin, Li},
	month = feb,
	year = {2020},
}

@article{wiler2016implementation,
  title={Implementation of a front-end split-flow model to promote performance in an urban academic emergency department},
  author={Wiler, Jennifer L and Ozkaynak, Mustafa and Bookman, Kelly and Koehler, April and Leeret, Robert and Chua-Tuan, Jenny and Ginde, Adit A and Zane, Richard},
  journal={The Joint Commission Journal on Quality and Patient Safety},
  volume={42},
  number={6},
  pages={271--AP4},
  year={2016},
  publisher={Elsevier}
}

@article{yu_sequential_2008,
	title = {Sequential effects: {Superstition} or rational behavior?},
	volume = {21},
	issn = {1049-5258},
	shorttitle = {Sequential effects},
	journal = {Advances in Neural Information Processing Systems},
	author = {Yu, Angela J. and Cohen, Jonathan D.},
	year = {2008},
	pmid = {26412953},
	pmcid = {PMC4580342},
	pages = {1873--1880},
}

@article{zayas2016dynamic,
  title={Dynamic control of a tandem system with abandonments},
  author={Zayas-Cab{\'a}n, Gabriel and Xie, Jingui and Green, Linda V and Lewis, Mark E},
  journal={Queueing Systems},
  volume={84},
  pages={279--293},
  year={2016},
  publisher={Springer}
}

@article{zhang_causal_nodate,
	title = {Causal {Inference} with {Non}-{IID} {Data} using {Linear} {Graphical} {Models}},
	language = {en},
	author = {Zhang, Chi and Mohan, Karthika and Pearl, Judea},
	pages = {12},
 year = {2022}
}
